\documentclass[5p, lefttitle, twocolumn]{elsarticle}

\usepackage{lineno,hyperref}
\modulolinenumbers[5]

\journal{Journal of \LaTeX\ Templates}

\usepackage{graphicx}
\usepackage{xcolor}
\usepackage{multirow}
\usepackage{rotating}
\usepackage{array}

\begin{document}

\begin{frontmatter}

\title{Deep Learning Based Brain Tumor Segmentation: A Survey}

\author[UoL]{Zhihua Liu}
\author[UoL]{Lei Tong}
\author[UoLancaster]{Zheheng Jiang}
\author[UoL]{Long Chen}
\author[UoL]{Feixiang Zhou}
\author[QMUL]{Qianni Zhang}
\author[Xidian]{Xiangrong Zhang}
\author[Bielefeld]{Yaochu~Jin}
\author[UoL]{Huiyu Zhou\corref{mycorrespondingauthor}}
\cortext[mycorrespondingauthor]{Corresponding author}
\ead{hz143@leicester.ac.uk}

\address[UoL]{School of Computing and Mathematical Sciences, University of Leicester, United Kingdom}
\address[UoLancaster]{School  of  Computing  and  Communications,  University  of Lancaster, United Kingdom}
\address[QMUL]{School of Electronic Engineering and Computer Science, Queen Mary, University of London, United Kingdom}
\address[Xidian]{School of Artificial Intelligence, Xidian University, China}
\address[Bielefeld]{Faculty of Technology, Bielefeld University, Germany}


\begin{abstract}
Brain tumor segmentation is one of the most challenging problems in medical image analysis. The goal of brain tumor segmentation is to generate accurate delineation of brain tumor regions. In recent years, deep learning methods have shown promising performance in solving various computer vision problems, such as image classification, object detection and semantic segmentation. A number of deep learning based methods have been applied to brain tumor segmentation and achieved promising results. Considering the remarkable breakthroughs made by state-of-the-art technologies, we use this survey to provide a comprehensive study of recently developed deep learning based brain tumor segmentation techniques. More than 100 scientific papers are selected and discussed in this survey, extensively covering technical aspects such as network architecture design, segmentation under imbalanced conditions, and multi-modality processes. We also provide insightful discussions for future development directions.
\end{abstract}


\end{frontmatter}


\begin{figure*}
\centering
  \includegraphics[width=0.8\textwidth]{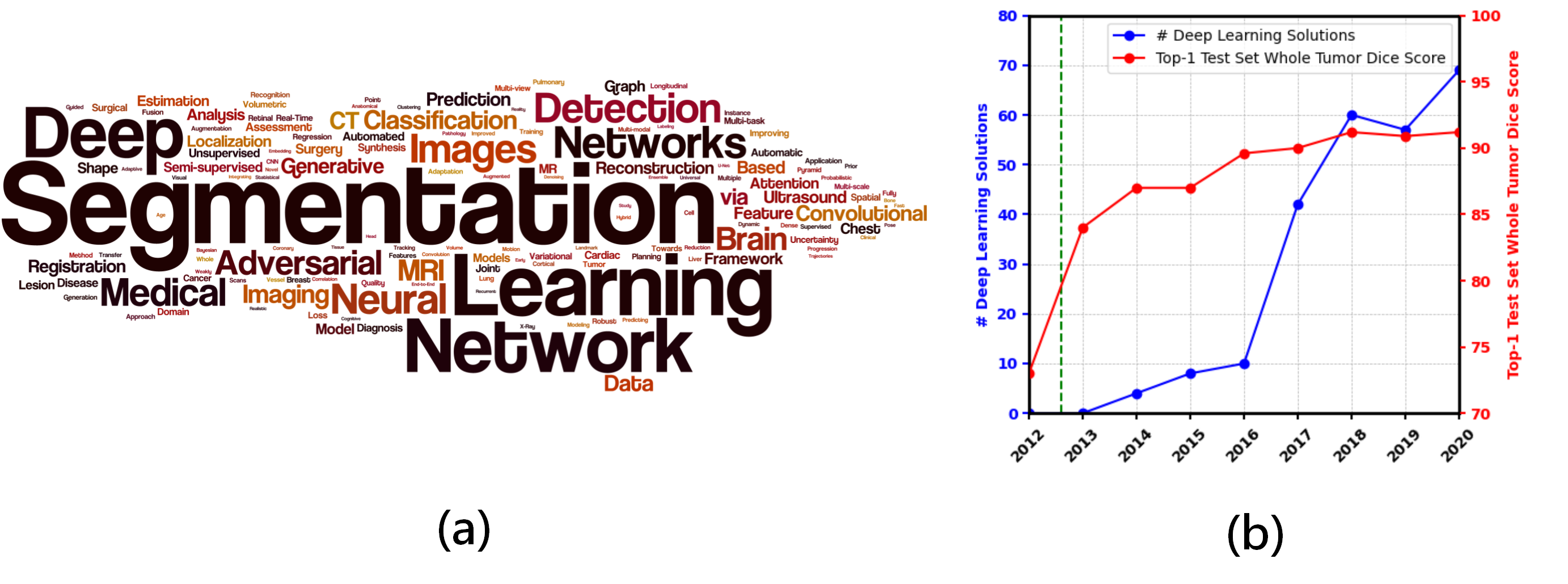}
\caption{Growth of scientific attention on deep learning based brain tumor segmentation. (a) Keyword frequency map in MICCAI from 2018 to 2020. The size of the keyword is proportional to the frequency of the word. We observe that 'brain', 'tumor', 'segmentation', and 'deep learning' have drawn large research interests in the community. (b) Blue line represents the number of deep learning based solutions in The Multimodal Brain Tumor Segmentation Challenge (BraTS) in each year. Red line represents the Top-1 whole tumor dice score of the test set in each year. Researchers shift their interests to deep learning based segmentation methods due to the powerful feature learning ability and systematic performance due to deep learning techniques since 2012 (green dashed line). Best viewed in colors.}
\label{fig:importance}       
\end{figure*}

\section{Introduction}
\label{intro}

Medical imaging analysis has been commonly involved in basic medical research and clinical treatment, e.g. computer-aided diagnosis \cite{doi2007computer}, medical record data management \cite{lavin1998system}, medical robots \cite{taylor2016medical} and image-based applications \cite{litjens2017survey}. Medical image analysis provides useful guidance for medical professionals to understand diseases and investigate clinical challenges in order to improve health-care quality. Among various tasks in medical image analysis, brain tumor segmentation has attracted much attention in the research community, which has been continuously studied (illustrated in Fig. \ref{fig:importance} (a)). In spite of tireless efforts of researchers, as a key challenge, accurate brain tumor segmentation still remains to be solved, due to various challenges such as location uncertainty, morphological uncertainty, low contrast imaging, annotation bias and data imbalance. With the promising performance made by powerful deep learning methods, a number of deep learning based methods have been applied upon brain tumor segmentation to extract feature representations automatically and achieve accurate and stable performance as illustrated in Fig. \ref{fig:importance} (b).

Glioma is one of the most primary brain tumors that stems from glial cells. World Health Organization (WHO) reports that glioma can be graded into four different levels based on microscopic images and tumor behaviors \cite{louis20162016}. Grade I and II are Low-Grade-Gliomas (LGGs) which are close to benign with slow-growing pace. Grade III and IV are High-Grade-Gliomas (HGGs) which are cancerous and aggressive. Magnetic Resonance Imaging (MRI) is one of the most common imaging methods used before and after surgery, aiming at providing fundamental information for the treatment plan.

\begin{figure}[hbt]
    \centering
    \includegraphics[width=0.48\textwidth]{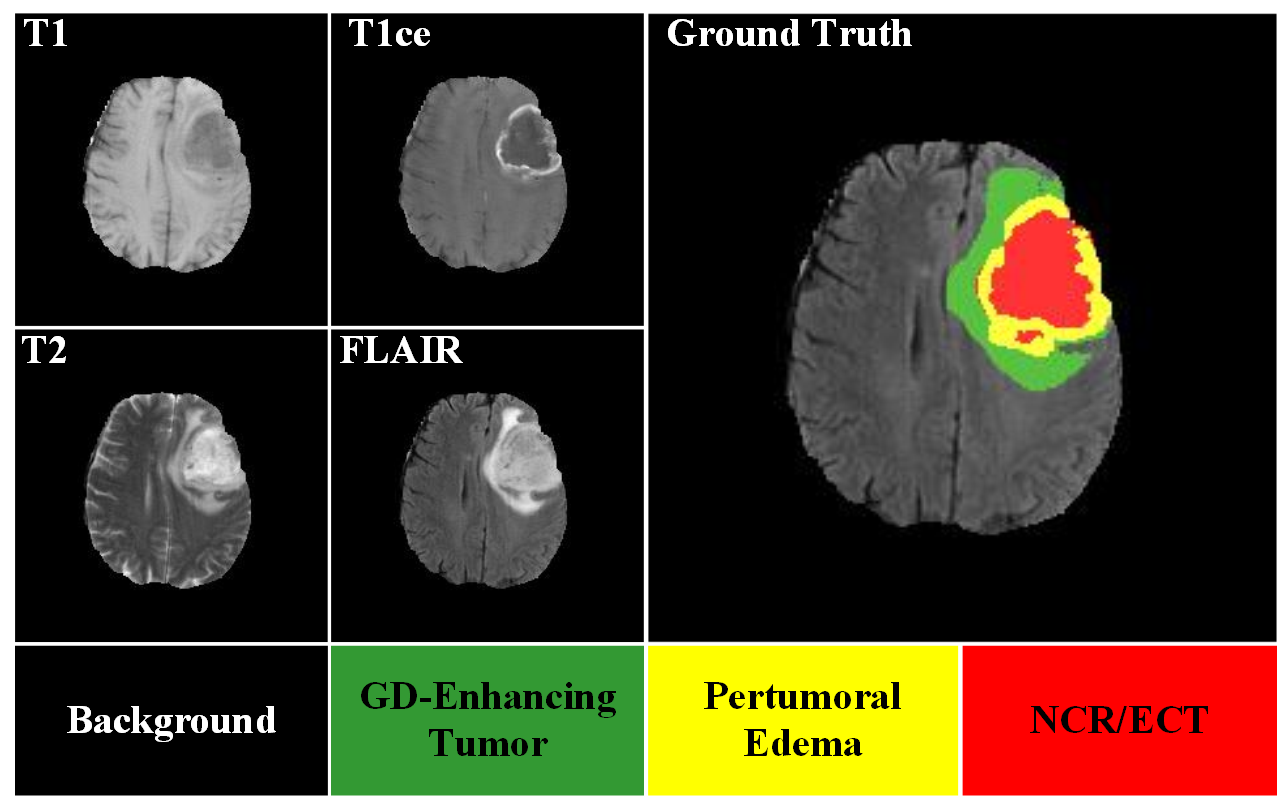}
    \caption{Exemplar input dataset with different MRI modalities and corresponding ground truth segmentation map. Each frame represents a unique MRI modality. The last frame on the right is the ground truth with corresponding manual segmentation annotation. Different colors represent different tumor sub-regions, i.e., gadolinium (GD) enhancing tumor (green), pertumoral edema (yellow) and necrotic and non-enhancing tumor core (NCR/ECT) (red). Best viewed in colors.}
    \label{fig:task}
\end{figure}

Image segmentation plays an active role in gliomas diagnosis and treatment. For example, an accurate glioma segmentation mask may help surgery planning, postoperative observations and improve the survival rate \cite{baid2021rsna}, \cite{bakas2017advancing}, \cite{menze2014multimodal}. To quantify the outcome of image segmentation, we define the task of \textit{brain tumor segmentation} as follows: Given an image from one or multiple image modality (e.g. multiple  MRI sequences), the system aims to automatically segment the tumor area from the normal tissues and to classify each voxel or pixel of the input data into a pre-set sub-region category. Finally, the system returns the segmentation map of the corresponding input. Fig. \ref{fig:task} shows one exemplar HGG case with different MRI sequences as input and corresponding ground truth segmentation map.

\subsection{Difference from Previous Surveys}
\label{sec: Difference from Previous Surveys}

A number of notable brain tumor segmentation surveys have been published in the last few years. We present recent relevant surveys with details and highlights in Table \ref{tab:survey}. Among them, the closest survey paper to ours is presented by Ghaffari et al.\cite{ghaffari2019automated}. The authors in \cite{ghaffari2019automated} covered a majority of submissions from BraTS2012 to BraTS2018 challenges, lacking, however, an analyses based on methodology category and highlights. Two recent surveys by Kapoor et al. \cite{kapoor2017survey} and Hameurlaine et al. \cite{hameurlaine2019survey} also focused on the summarisation of classic brain tumor segmentation methods. However, both of them lacked the technical analysis and discussion of deep learning based segmentation methods. A survey of early state-of-the-art brain tumor segmentation methods before 2013 was presented in \cite{gordillo2013state}, where most of the proposals before 2013 combined conventional machine learning models with hand-crafted features. Liu et al. \cite{liu2014survey} reported a survey on MRI based brain tumor segmentation in 2014. This survey does not include deep learning based methods as well. Nalepa et al. \cite{nalepa2019data} analysed the technical details and impacts of different kinds of data augmentation methods with the application to brain tumor segmentation, while ours focuses on the technical analysis of deep learning based brain tumor segmentation methods.

\begin{table*}[]
\centering
\caption{A summary of the existing surveys relates to the topic 'brain tumor segmentation'.}
\label{tab:survey}
\begin{tabular}{|p{0.30\textwidth}|p{0.30\textwidth}|p{0.03\textwidth}|p{0.30\textwidth}|}
\hline
\textbf{Survey Title}                                                                                                                           & \textbf{Venue}                                                                                        & \textbf{Year} & \textbf{Remarks}                                                                                                                                      \\ \hline
Automated brain tumor segmentation using multimodal brain scans: a survey based on models submitted to the BraTS 2012--2018 challenges \cite{ghaffari2019automated} & IEEE Reviews in Biomedical Engineering                                                       & 2019 & A review of challenge submissions of BraTS during 2012-2018.                                                                                   \\ \hline
A survey on brain tumor detection using image processing techniques \cite{kapoor2017survey}                                                                    & 2017 7th International Conference on Cloud computing, Data science \& Engineering-confluence & 2017 & A review of general brain tumor segmentation methods.                                                                                        \\ \hline
Survey of brain tumor segmentation techniques on magnetic resonance imaging \cite{hameurlaine2019survey}                                                            & Nano Biomedicine and Engineering                                                             & 2019 & A general summary of classic brain tumor segmentation methods.                                                                               \\ \hline
State of the art survey on MRI brain tumor segmentation \cite{gordillo2013state}                                                                               & Magnetic Resonance Imaging                                                                   & 2013 & Review on convolutional neural networks used for brain MRI image analysis.                                                                   \\ \hline
A survey of MRI-based brain tumor segmentation methods \cite{liu2014survey}                                                                                & Tsinghua Science and Technology                                                              & 2014 & Review on MRI based brain tumor segmentation methods.                                                                                        \\ \hline
Data augmentation for brain-tumor segmentation: a review \cite{nalepa2019data}                                                                              & Frontiers in Computational Neuroscience                                                      & 2019 & Analysed the technical details and impacts of different kinds of data augmentation methods with the application to brain tumor segmentation. \\ \hline
A survey on deep learning in medical image analysis \cite{litjens2017survey}                                                                                   & Medical Image Analysis                                                                       & 2017 & A comprehensive review on deep learning based medical image analysis.                                                                        \\ \hline
Deep convolutional neural networks for brain image analysis on magnetic resonance imaging: a review \cite{bernal2019deep}                                   & Artificial Intelligence in Medicine                                                          & 2018 & A  review on use  of  deep convolutional neural networks for brain image analysis.                                                           \\ \hline
Deep learning for brain MRI segmentation: state of the art and future directions \cite{akkus2017deep}                                                      & Journal of Digital Imaging                                                                   & 2017 & A survey on deep learning for brain MRI segmentation.                                                                                        \\ \hline
A guide to deep learning in healthcare \cite{esteva2019guide}                                                                                                & Nature Medicine                                                                              & 2019 & A survey on deep learning for health-care.                                                                                                   \\ \hline
Deep learning for generic object detection: A survey    \cite{liu2020deep}                                                                               & International Journal of Computer Vision                                                     & 2020 & A  comprehensive  review  on  deep  learning based object detection.                                                                         \\ \hline
Deep learning \cite{lecun2015deep}                                                                                                                         & Nature                                                                                       & 2015 & An introduction review on deep learning and its application.                                                                                 \\ \hline
Recent advances in convolutional neural networks \cite{gu2018recent}                                                                                       & Pattern Recognition                                                                          & 2018 & A survey on convolutional neural networksand its application on computer vision, language processing and speech.                             \\ \hline
Deep   Learning   Based   Brain   Tumor Segmentation: A Survey                                                                         & Ours                                                                                         & -    & A  comprehensive  survey  of  deep  learning based brain tumor segmentation.                                                                 \\ \hline
\end{tabular}
\end{table*}

There is a number of representative survey papers published with similar topics in recent years. Litjens et al.  \cite{litjens2017survey} summarised recent medical image analysis applications with deep learning techniques. This survey gives an over of broad studies on medical image analysis including several state-of-the-art deep learning based brain tumor segmentation methods before 2017. Bernal et al. \cite{bernal2018deep} reported a review focusing on the use of deep convolutional neural networks for brain image analysis. This review only highlights the application of deep convolutional neural networks. Other important learning strategies such as segmentation under imbalance condition and learning from multi-modality were not mentioned. Akkus et al. \cite{akkus2017deep} presented a survey on deep learning for brain MRI segmentation. Recently, Esteva et al. \cite{Esteva2019} presented a survey on deep learning for health-care applications. This survey summarized how deep learning in computer vision, natural language processing, reinforcement learning and generalized methods promote health-care applications. For a broader view of object detection and semantic segmentation, a survey was recently published in \cite{liu2020deep}, providing the implications on object detection and semantic segmentation.

\begin{figure}
    \centering
    \includegraphics[width=0.5\textwidth]{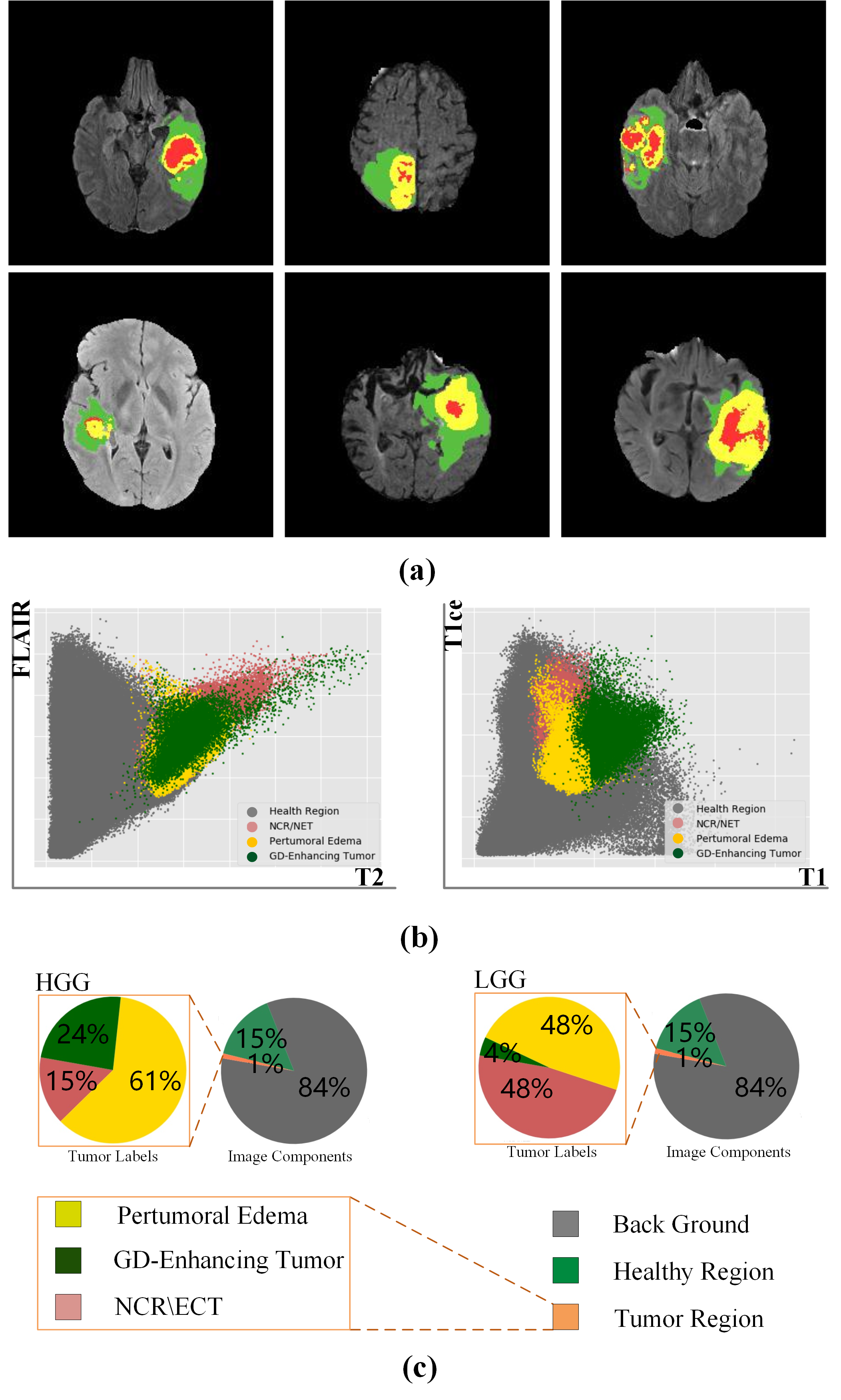}
    \caption{Challenges in segmentation of brain glioma tumors. (a) shows glioma tumor exemplars with various sizes and locations inside the brain. (b) and (c) show the statistical information of the training set in the multimodal brain tumor segmentation challenge 2017 (BraTS2017). The left hand side of (b) shows the FLAIR and T2 intensity projection, and the right hand side shows the T1ce and T1 intensity projection. (c) is the pie chart of the training data with labels, where the top figure shows the HGG labels while the bottom figure shows the LGG labels. We here experience region and label imbalance problems. Best viewed in colors.}
    \label{fig:challenge}
\end{figure}

Narrowly speaking, the word "deep learning" means using neural network models with stacked functional layers (usually the layer number $>$ 5) \cite{goodfellow2016deep}. Neural networks are able to learn high dimensional hierarchical features and approximate any continuous functions \cite{lin2018resnet}, \cite{yarotsky2017error}. Considering the achievements and recent advances of deep neural networks, several surveys have reported the developed deep learning techniques, such as \cite{gu2018recent} and \cite{lecun2015deep}.

\subsection{Scope of This Survey}
\label{sec:Scope of This Survey}

In this survey, we have collected and summarized the research studies reported on over one hundred scientific papers. We have examined major journals in the scientific community such as Medical Image Analysis and IEEE Transactions on Medical Imaging. We also evaluated proceedings of major conferences, such as ISBI, MICCAI, IPMI, MIDL, CVPR, ECCV and ICCV, to retain frontier medical imaging research outcomes. We reviewed annual challenges and their related competition entries such as The Multimodal Brain Tumor Segmentation Challenge (BraTS). In addition, the pre-printed versions of the established methods on arXiv are also included as a source of information.

The goal of this survey is to present a comprehensive technical review of deep learning based brain tumor segmentation methods, according to architectural categories and strategy comparisons. We wish to explore how different architectures affect the segmentation performance of deep neural networks and how different learning strategies can be further improved for various challenges in brain tumor segmentation. We cover diverse high level perspectives, including effective architecture design, imbalance segmentation and multi-modality process. The taxonomy of this survey is made (Fig. \ref{fig:taxonomy}) such that our categorization can help the reader to understand the technical similarities and differences between segmentation methods. The proposed taxonomy may also enable the reader to identify open challenges and future research directions.

We first present the background information of deep learning based brain tumor segmentation methods in Section \ref{sec:Background} and the rest of this survey is organised as follows: In Section \ref{sec:Designing Effective Segmentation Network}, we review the design paradigm of effective segmentation modules and network architectures. In Section \ref{sec:Segmentation under Imbalanced Condition}, we categorise, explore and compare the solutions for tackling the data imbalance issue, which is a long-standing problem in brain tumor segmentation. As multi-modality provides promising solutions towards accurate brain tumor segmentation, we finally review the methods of utilising multi-modality information in Section \ref{sec:Utilizing Multi Modality Information}. We conclude this paper in Section \ref{sec:Conclusion}. We also build up a regularly maintained project page to accommodate the updates related to this survey.\footnote{http://github.com/ZhihuaLiuEd/SoTA-Brain-Tumor-Segmentation.}

\begin{figure*}[hbt]
    \centering
    \includegraphics[width=\textwidth]{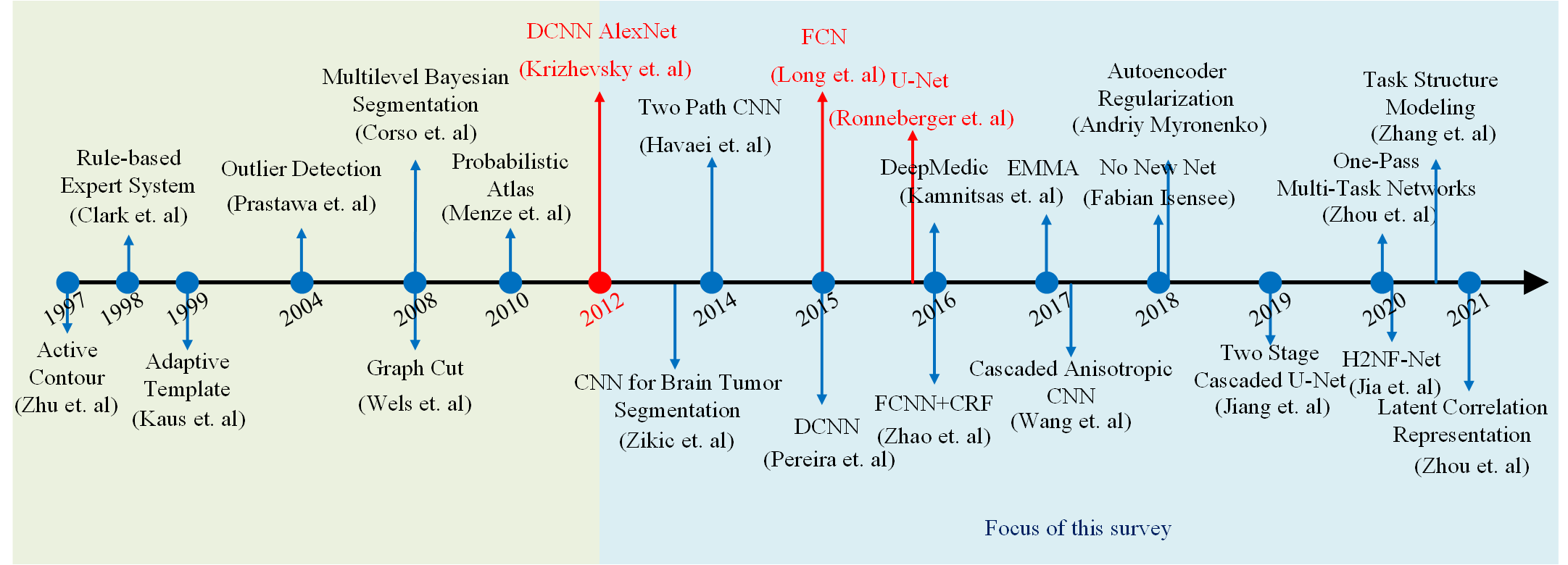}
    \caption{The evolution of brain tumor segmentation with selective milestones over the past decade. Best viewed in colors.}
    \label{fig:progresssummary}
\end{figure*}

\section{Background}
\label{sec:Background}

\subsection{Research Challenges}
\label{sec:Research Challenges}
Despite significant progress that has been made in brain tumor segmentation, state-of-the-art deep learning methods still experience difficulties with several challenges to be solved. The challenges associated with brain tumor segmentation can be categorised as follows:

\begin{enumerate}
     \item{\textbf{Location Uncertainty}}
    Glioma is mutated from gluey cells which surround nerve cells. Due to the wide spatial distribution of gluey cells, either High-Grade Glioma (HGG) or Low-Grade Glioma (LGG) may appear at any location inside the brain. 
    \item{\textbf{Morphological Uncertainty}} Different from a rigid object, the morphology, e.g. shape and size, of different brain tumors varies with large uncertainty. As the external layer of a brain tumor, edema tissues show different fluid structures, which barely provide any prior information for describing the tumor's shapes. The sub-regions of a tumor may also vary in shape and size.
    \item{\textbf{Low Contrast}} High resolution and high contrast images are expected to contain diverse image information \cite{liu2016ssd}. Due to the image projection and tomography process, MRI images may be of low quality and low contrast. The boundary between biological tissues tends to be blurred and hard to detect. Cells near the boundary are hard to be classified, which makes precise segmentation more difficult and harder to achieve.
    \item{\textbf{Annotation Bias}} Manual annotation highly depends on individual experience, which can introduce an annotation bias during data labeling. As shown in Fig. \ref{fig:challenge} (a), it seems that some annotations tend to connect all the small regions together while the other annotations can label individual voxels precisely. The annotation biases have a huge impact on the segmentation algorithm during the learning process \cite{chen2021understanding}.
    \item{\textbf{Imbalanced Issue}} As shown in Fig. \ref{fig:challenge} (b) and  (c), there exists an imbalanced number of voxels in different tumor regions. For example, the necrotic/non-enhancing tumor core (NCR/ECT) region is much smaller than the other two regions. The imbalanced issue affects the data-driven learning algorithm as the extracted features may be highly influenced by large tumor regions \cite{bulo2017loss}.
\end{enumerate}

\subsection{Progress in the Past Decades}
\label{sec:Progress in the Past Decades}
Representative research milestones of brain tumor segmentation are shown in Fig. \ref{fig:progresssummary}. In the late 90s', researchers Zhu et al. \cite{zhu1997computerized} started to use a Hopfield Neural Network with active contours to extract the tumor boundary and dilate the tumor region. However, training a neural network was highly constrained due to the computational resource limitation and technical supporting. From late 90s' to early 20s', most of the brain tumor segmentation methods focused on traditional machine learning algorithms with hand-crafted features, such as expert systems with multi-spectral histogram \cite{clark1998automatic}, segmentation with templates \cite{kaus1999segmentation}, \cite{prastawa2004brain}, graphical models with intensity histograms \cite{corso2008efficient}, \cite{wels2008discriminative}, tumor boundary detection from latent atlas \cite{menze2010generative}. These early works pioneered the use of machine learning in solving brain tumor segmentation problems. However, early research works have significant shortcomings. First, most of the early works only focused on the segmentation of the whole tumor region, that is, the segmentation result has only one category. Compared with recent brain tumor segmentation algorithms, early works are formulated with strong conditions, relying on unrealistic assumptions. Second, manually designed feature engineering is constrained by prior knowledge, which cannot be fully generalised. Last but not least, early research works fail to address some challenges such as appearance uncertainty and data imbalance.

\begin{figure*}[hbt]
    \centering
    \includegraphics[width=\textwidth]{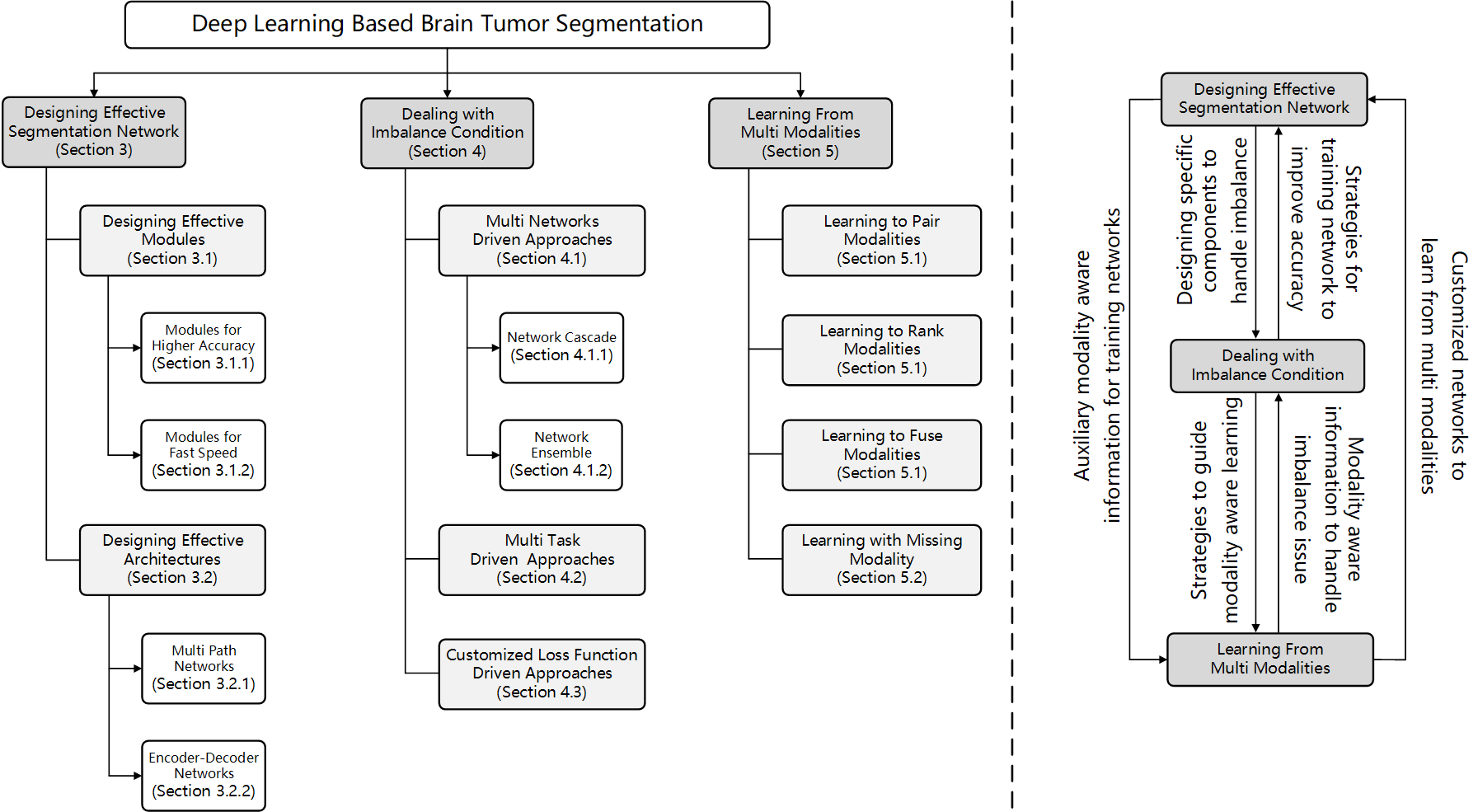}
    \caption{Our proposed taxonomy of deep learning based brain tumor segmentation methods. Best viewed in colors.}
    \label{fig:taxonomy}
\end{figure*}

With the revolutionary breakthrough by deep learning technology \cite{krizhevsky2012imagenet}, researchers began to focus on using deep neural networks to solve various practical problems. Pioneering works from Zikic et al.\cite{zikic2014segmentation}, Havaei et al.\cite{havaei2017brain}, Pereira et al.\cite{pereira2016brain} intend to design customized deep convolutional neural networks to achieve accurate brain tumor segmentation. With breakthrough brought by Fully Convolutional Network (FCN) \cite{long2015fully} and U-Net \cite{ronneberger2015u}, recent innovations \cite{isensee2018no}, \cite{zhao2018deep} on brain tumor segmentation focus on building encoder-decoder networks without fully connected layers to achieve end-to-end tumor segmentation.

A long-standing challenge in brain tumor segmentation is data imbalance. To effectively deal with the imbalance problem, researchers try different solutions, such as network cascade and ensemble \cite{jiang2019two}, \cite{kamnitsas2017ensembles}, \cite{wang2017automatic},  multi-task learning \cite{myronenko20183d}, \cite{zhou2020one}, and customized loss functions \cite{sudre2017generalised}. Another solution is to fully utilise information from multi-modality. Recent research focused on modality fusion \cite{zhang2021cross} and dealing with modality missing \cite{9399263}.

Based on the evolution, we generally categorise the existing deep learning based brain tumor segmentation methods into three categories, i.e., methods with effective architectures, methods for dealing with imbalanced condition and methods of utilising multi-modality information. Fig. \ref{fig:taxonomy} shows a taxonomy of the research work in deep learning based brain tumor segmentation.

\subsection{Related Problems}
\label{sec:Related Problems}
There are a number of unsolved problems that relates to brain tumor segmentation. \textit{Brain tissue segmentation} or \textit{anatomical brain segmentation} aims to label each unit with a unique brain tissue class. Their task assumes that the brain image does not contain any tumor tissue or other anomalies \cite{de2015deep}, \cite{patenaude2011bayesian}. The goal of \textit{white matter lesion segmentation} is to segment the white matter lesion from the normal tissue. In their task, the white matter lesion does not contain sub-regions such as tumor cores, where segmentation may be achieved through binary classification methods. \textit{Tumor detection} aims to detect abnormal tumors or lesion and reports the predicted class of each tissue. Generally, this task has the bounding box as the detection result and the label as the classification result \cite{dou2015automatic}, \cite{dou2016automatic},\cite{ghafoorian2017deep}. It is worth mentioning that some research methods in \textit{brain tumor segmentation} only return the single label segmentation mask or the center point of the tumor core without sub-region segmentation. In our paper, we focus on tumor segmentation with sub-region level semantic segmentation as the main topic. \textit{Disorder classification} is to extract pre-defined features from brain scan images and then classify feature representations into graded disorders such as High-Grade-Gliomas (HGGs) vs Low-Grade-Gliomas (LGGs), Mild Cognitive Impairment (MCI) \cite{suk2016state}, Alzheimer’s Disease (AD) \cite{suk2016deep} and Schizophrenia \cite{pinaya2016using}. \textit{Survival Prediction} identifies tumors' patterns and activities \cite{yoo2016deep} in order to predict the survival rate as a supplementary to clinical diagnosis \cite{van2017deep}. Both disorder classification and survival prediction can be regarded as down-stream tasks, based on the tumor segmentation outcomes.

\subsection{Contributions of this survey}
\label{sec: Contribution of this survey}
A large number of deep learning based brain tumor segmentation methods have been published with promising results. Our paper, as a platform, provides a comprehensive and critical survey of state-of-the-art  brain tumor segmentation methods. We anticipate that this survey supplies useful guidelines and coherent technical insights to academia and industry. The major contributions of this survey can be summarised as follows: 
\begin{enumerate}
    \item To  our  best  knowledge,  this  is  the  first  survey  to catergorise and outline deep  learning based  brain  tumor  segmentation  methods  with  a structured  taxonomy  of  various  important  technical perspectives.
    \item We present the reader with a summarisation of technological progress of deep learning base brain tumor segmentation  with  detailed  background  information  and system comparisons (e.g. Tables \ref{tab:survey}, \ref{tab:opensource}).
    \item We carefully and extensively compares existing methods based on results from public accessible challenges and datasets (e.g. Tables \ref{table:effectivenetwork}, \ref{table:imbalancesegmentation}, \ref{table:multimodality}.), with critical summaries and insightful discussions.
\end{enumerate}

\section{Designing Effective Segmentation Networks}
\label{sec:Designing Effective Segmentation Network}

Compared with complex feature engineering pipelines to extract useful features, recent deep learning mainly relies on designing effective deep neural networks to automatically extract high-dimensional discriminative features. Designing effective modules and network architectures has become one of the important factors for achieving accurate segmentation performance. In this section, we reviewed two important design guidelines for deep learning based brain tumor segmentation: to design effective modules and network architecture. 

There are mainly two principles to follow when designing effective components. One is to learn high level semantics and localise precious targets, through the enlargement of the receptive field \cite{li2017brain}, \cite{lopez2017dilated}, \cite{zhao2017automatic}, attention mechanism \cite{islam2019brain}, \cite{wang2019global}, \cite{zhou2020one} feature fusion update \cite{liu2020brain}, \cite{zhou2020multi} and other forms. The other way is to reduce the amount of the network parameters and speed up during training and inference, thereby saving computational time and resources\cite{andermatt2017multi}, \cite{brugger2019partially}, \cite{chen20193d}, \cite{cheng2019memory}, \cite{pendse2020memory}, \cite{zhao2016multiscale}, \cite{shen2017efficient}.

\begin{figure*}
\centering
  \includegraphics[width=0.83\textwidth, height=22cm]{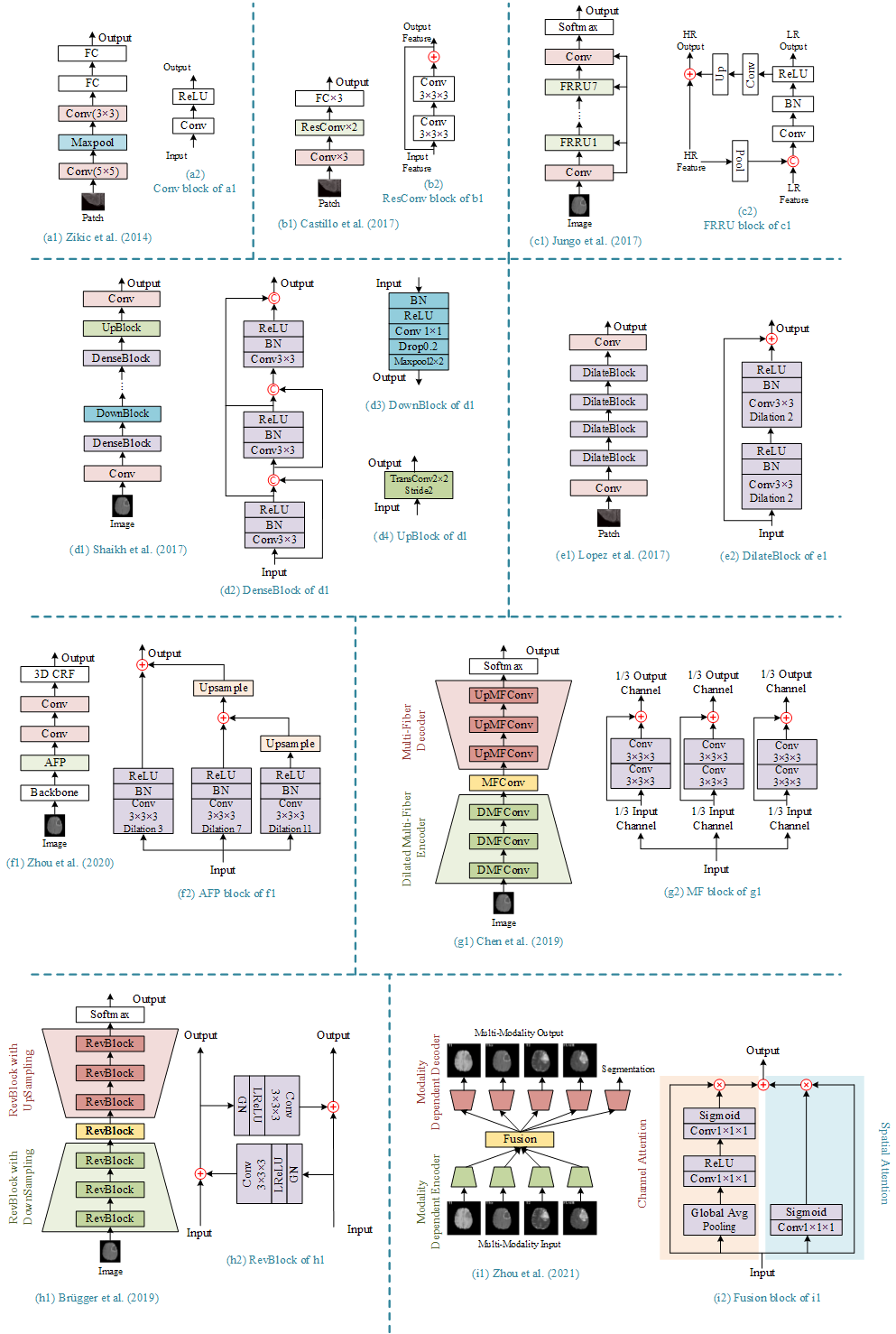}
\caption{Structural comparison between representative methods based on designing effective network modules and architectures. From top-left to bottom-right: (a1) CNN in \cite{zikic2014segmentation}, (b1) CNN with (b2) residual convolution module \cite{castillo2017volumetric}, (c1) CNN with (c2) full resolution residual unit \cite{jungo2017towards}, (d1) CNN with (d2) dense connection module \cite{shaikh2017brain}, (e1) CNN with (e2) residual dilation block \cite{lopez2017dilated}, (f1) CNN with (f2) atrous convolution feature pyramid module \cite{zhou2020afpnet}, (g1) FCN with (g2) multi-fiber unit \cite{chen20193d}, (h1) FCN with (h2) reversible block \cite{brugger2019partially} and (i1) FCN with (i2) modality fusion module \cite{zhou2021latent}. Best viewed in colors.}
\label{fig:modulecomp}       
\end{figure*}

The design of the network architecture is mainly reflected in the transition from a single-channel network to a multi-channel network, from a network with fully connected layers to a fully convolutional network, from a simple network to a deep cascaded network. The purpose is to deepen the network, enhance the feature learning ability of the network and completes more precise segmentation. In the following, we divide theses methods and review them comprehensively. A systematical comparison between various network architectures and modules is shown in Fig. \ref{fig:modulecomp}.

\subsection{Designing Specialised Modules}
\subsubsection{Modules for Higher Accuracy}
Numerous methods for brain tumor segmentation focuses on designing effective modules inside neural networks, aiming to stabilise training, learning informative, discriminative, and conducive features for accurate segmentation. Early design work followed the pattern of well-known networks such as AlexNet \cite{krizhevsky2012imagenet} and gradually deepened the network depth by stacking convolutional blocks. Early research works such as  \cite{dvovrak2015local}, \cite{rao2015brain} and \cite{zhao2018deep} stacked several blocks with convolutional layers composed of a large kernel size (typically greater than 5), pooling layers and activation layers together. Blocks with a large size convolution kernel enable us to capture useful details with a large number of parameters to be trained. Other research works such as \cite{zikic2014segmentation} and \cite{pereira2016brain} followed the pattern of VGG \cite{simonyan2014very} to build convolutional layers with a small sized kernel (typically 3) as basic blocks. Further research works such as \cite{havaei2017brain} stacked hybrid blocks with a combination of different kernel sizes, where large sized kernels tend to find global features (such as tumor location and size) with a large receptive field and small kernels tend to contain local features (such as boundary and texture) with a small receptive field. As stacking two $3\times3$ convolutional layers leads to equal sized reception fields while maintaining less parameters, compared with a single $5\times5$ layer, most recent tumor segmentation works constructed basic network blocks, based on stacking $3\times3$ layers, and started to extend to volumetric reconstruction in MRI with $3\times3\times3$ kernels \cite{casamitjana20163d}, \cite{jesson2017brain}.

As the number of stacked layers increases, the network is getting deeper, causing the issue of gradient explosion and vanishing during the training process. In order to stabilise system training and reach higher segmentation accuracy, early brain tumor segmentation methods such as \cite{chang2016fully} and \cite{castillo2017volumetric} followed ResNet\cite{he2016deep} and introduced residual connection into module design. Residual connection helps solving the problem of gradient vanishing and explosion, by adding the input of a convolution module to its output, which avoids degradation and converges faster with better accuracy. Now, residual connection has become one of the standard operations for designing modules and complex network architectures. In the following works \cite{ghaffari2020brain}, \cite{shaikh2017brain}, \cite{wang2020modality} and \cite{DBLP:journals/cbm/ZhouHSDC20}, the authors followed Densenet \cite{huang2017densely} and expanded residual connection to dense connection. Although dense connection design looks more conducive to gradient back-propagation, the complex close connection structure can cause multiple usage of the computing memory during the network training.

By stacking convolution modules and using residual connections inside and outside modules, neural networks can be deeper and features can be learnt with higher dimensions and uncertainty. However, this process may lead to the sacrifice of spatial resolution, whereas the resolution of high dimensional feature maps is much smaller than that of the original data. In order to preserve the spatial resolution of data whilst still expanding the receptive field, \cite{li2017brain}, \cite{lopez2017dilated}, \cite{zhao2017automatic} replaced the standard convolution layer with a dilated convolution layer \cite{yu2017dilated}. The dilated convolution comes up with several benefits. First, dilation convolution enlarges the receptive field without introducing additional parameters. Larger receptive fields are helpful for segmenting large-area targets, such as edema. Second, dilated convolution avoids the loss of spatial resolution. Thus, the position of the object to be segmented can be accurately localised in the original input space. However, the problem of incorrect localisation and segmentation of small structures remains to be solved. In response to this problem, \cite{choudhury2018segmentation} proposed to design a multi-scale dilation convolution or atrous spatial pyramid pooling module, capturing the semantic context that describes subtle details of the object.

\subsubsection{Modules for Efficient Computation}
Designing and stacking complex modules may help effectively learn high-dimensional discriminative features and achieve precise segmentation, but it requires high computational resources and long training and inference time. In response to this request, many works have adopted lightweight ideas in module design. With similar accuracy, fewer computing resources are required by lightweight architectures, training and reasoning time is shorter, and the speed is faster. \cite{andermatt2016multi} is one of the earliest research works aiming at speeding up brain tumor segmentation. The authors of \cite{andermatt2016multi} reordered the input data (a data sample rotated by 6-degrees) so that the samples with high visual similarity are placed closer in the memory, in an attempt to speed up I/O communication. Instead of managing the input data, \cite{cheng2019memory} chose to build a U-Net variant with decreased down-sampling channels to reduce the computational cost.

The above-mentioned works used less computational resources, but lose learning information and decreased segmentation accuracy. Inspired by reversible residual network \cite{gomez2017reversible}, \cite{brugger2019partially} introduced reversible blocks into U-Net where each layer's activation can be collected from the previous layer's output during the backward pass process. Thus, no additional memory is used to store intermediate activation and hence reduce memory cost. \cite{pendse2020memory} further extend reversible blocks by introducing Mobile Reversible Convolution Blocks (MBConvBlock) used in MobileNetV2 \cite{sandler2018mobilenetv2} and EfficientNet \cite{tan2019efficientnet}. In addition to the reversible computation design, MBConvBlock replaced standard convolutions with depthwise separable convolutions. Depthwise separable convolutions first split the computation of feature maps accordingly using depthwise convolution and merge the feature maps together using $1\times1\times1$ pointwise convolutions, which further reduced parameters compared with the standard convolution. Later research works, including 3DESPNet\cite{nuechterlein20183d} and DMFNet \cite{chen20193d}, further extend this idea with dilated convolutions, requiring less computational resources while preserving most spatial resolutions.

\subsection{Designing Effective Architectures}
A major factor that promotes prosperity and development of deep neural networks in various fields is to invest efforts in designing intelligent and effective network architectures. We divide most deep learning based brain tumor segmentation networks into single/multiple path networks and encoder-decoder networks according to the characteristics of network structures. Single and multiple path networks are used to extract features and classify the center pixels of the input patch. Encoder-Decoder networks are designed in an end-to-end fashion, that is, the encoder enables deep feature to be extracted from part of or the entire image, and then the decoder conducts feature-to-segmentation mapping. In the following subsections, we conduct a systematic analysis and comparison of variant architecture designs.

\subsubsection{Multi-Path Architecture}
\label{subsubsection:Multi-Path Architecture}

\begin{figure}[t]
    \centering
    \includegraphics[width=0.5\textwidth]{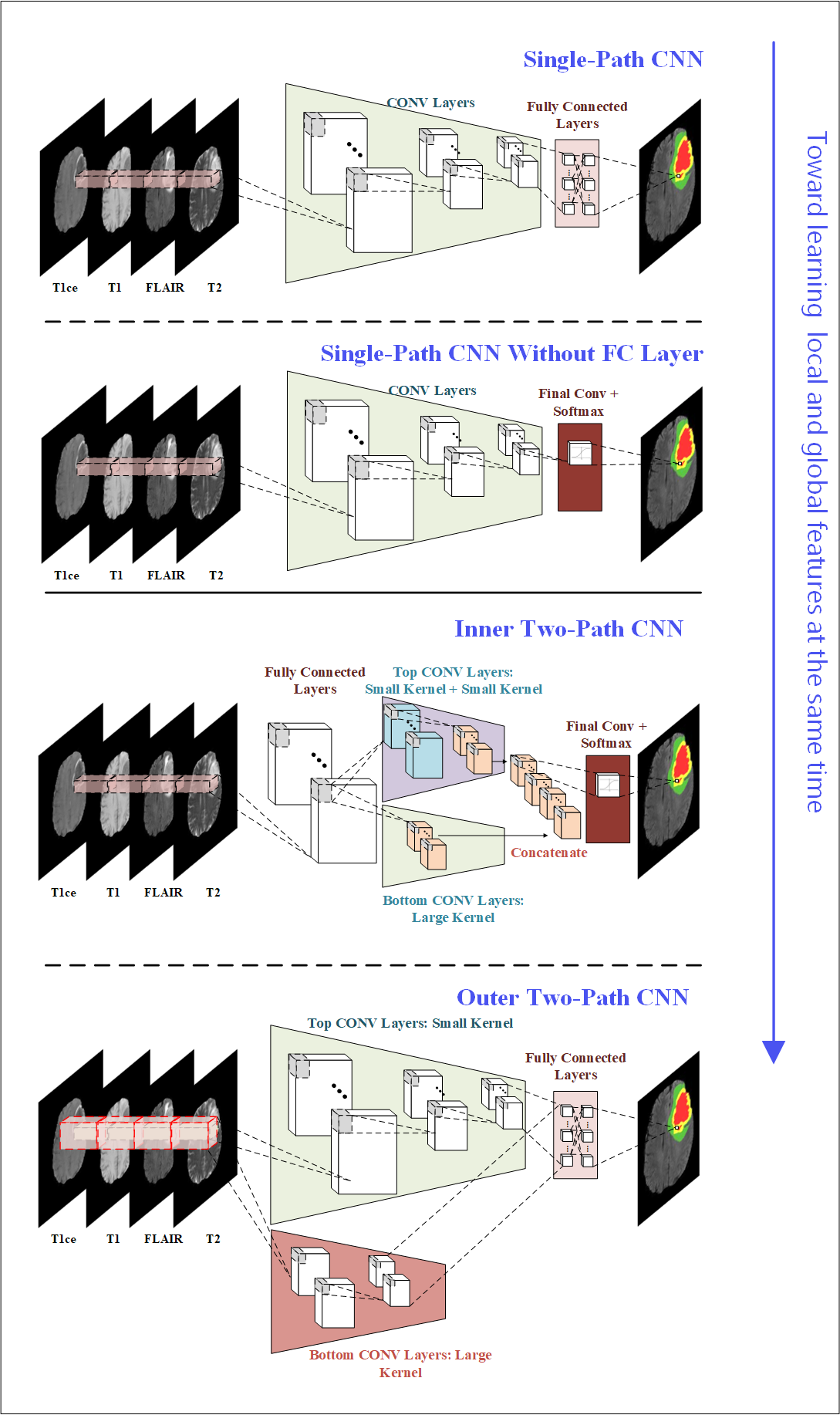}
    \caption{A high level comparison between single-path  and two-path CNN. Best viewed in colors.}
    \label{fig:pathnetwork}
\end{figure}

Here we refer network path as the flow of data processing (Fig. \ref{fig:pathnetwork}). Many research works e.g. \cite{pereira2016brain}, \cite{urban2014multi}, \cite{zikic2014segmentation} use single path networks due to their computational efficiency. Compared with single path networks, multi-path networks can extract different features from different pathways of different scales. The extracted features are combined (added or concatenated) together for further processing. A common interpretation is that a large scale path (path with a large size's kernel or input etc.) allows us to learn global features. Small scale's paths (paths with a small size's kernel or input etc.) allows us to learn features known as local features. Similar to the functionality mentioned in the previous section, global features tend to provide global information such as tumor location, size and shape while local features provide descriptive details such as tumor texture and boundary.

The work of Havaei et al. \cite{havaei2017brain} is one of the early multi-path network based solutions. The author reported a novel two pathway structure that learns local tumor information as well as global contexts. The local pathway uses a $7\times7$ convolution kernel and the global pathway uses a $13\times13$ convolution kernel. In order to utilise CNN architectures, the authors designed several variant architectures that concatenate CNN outputs. Castillo et al. \cite{castillo2017volumetric} used a 3 pathway CNN to segment brain tumors. Different from \cite{havaei2017brain} that used kernels in different scales, \cite{castillo2017volumetric} inputs each path with different sizes' patches e.g. patches with low ($15\times15$), medium($17\times17$) and normal ($27\times27$) resolutions. Thus, each path can learn specific features under the condition of different spatial resolutions. Inspired by \cite{havaei2017brain}, Akil et al. \cite{akil2020fully} extended the network structure with overlapping patch prediction methods, where the center of the target patch is associated with the neighbouring overlapping patches.

Instead of building multi-path networks with different sizes' kernels, other research works attempt to learn local-to-global information from the input directly.
For example, Kamnitsas et al. \cite{kamnitsas2017efficient} presented a dual pathway network which considers the input with different sizes' patches, known as the normal resolution input of size $25 \times 25 \times 25$ and the low resolution input of size $19 \times 19 \times 19$. Different from \cite{castillo2017volumetric}, the authors in \cite{kamnitsas2017efficient} applied small convolution kernels with a size of $3 \times 3 \times 3$ on both pathways. Later research works by Zhao et al. \cite{zhao2016multiscale} also designed a multi-scale CNN with a large scale path with the input size of $48 \times 48$, a middle scale path with the input size of $18 \times 18$ and a small scale path with the input size of $12 \times 12$.

\subsubsection{Encoder-Decoder Architecture}

\begin{figure*}[t]
    \centering
    \includegraphics[width=\textwidth]{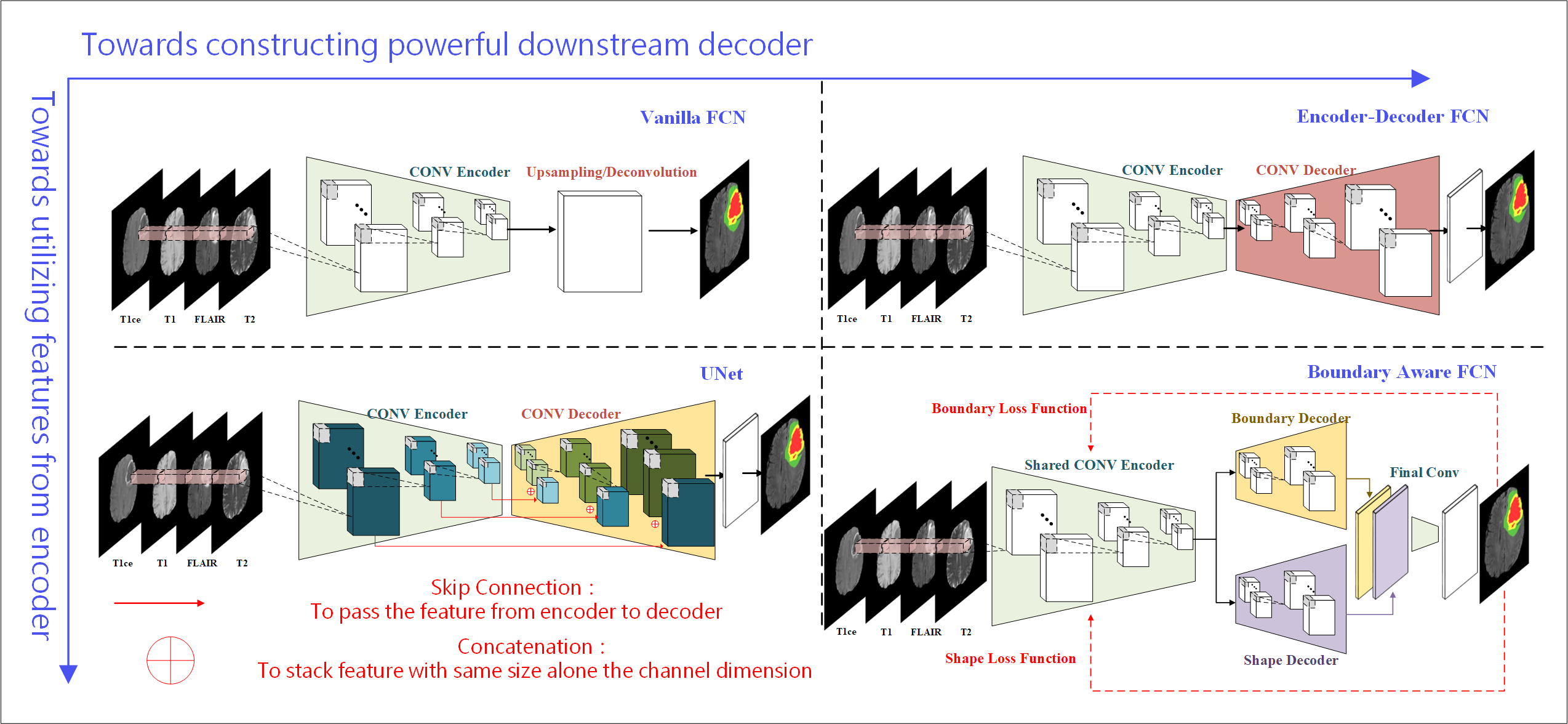}
    \caption{A high level comparison between different fully convolutional networks (FCNs). Best viewed in colors.}
    \label{fig:fcncompare}
\end{figure*}

The input of the single and multiple path network for brain tumor segmentation is a patch or a certain area of the image, and the output is the classification outcome of the patch or the classification outcome of the central pixel of the input. It is very challenging to promote an accurate mapping from the patch level to the category label. First of all, the segmentation performance of single and multiple path network is easily affected by the size and quality of the input patch. A small sized input patch holds incomplete spatial information, while a large sized patch requires more computational resources. Secondly, the feature-to-label mapping is mostly conducted by the last fully connected layer. A simple fully connected layer cannot fully represent the feature space where complicated fully connected layers may overload the computer's memory. Last but not least, this feature-to-label mapping is not of an end-to-end mode, which significantly increases the optimisation cost. To tackle these problems, recent research works start to use fully convolutional network (FCN) \cite{long2015fully} and U-Net \cite{ronneberger2015u} based encoder-decoder networks, establish an end-to-end fashion from the input image to the output segmentation map, and further improve the segmentation performance of networks.

Jesson et al. \cite{jesson2017brain} extended standard FCN by using a multi-scale loss function. One limitation of FCN is that FCN does not explicitly model the contexts in the label domain. In \cite{jesson2017brain}, the FCN variant minimised the multi-scale loss by combining higher and lower resolution feature maps to model the contexts in both image and label domains. In \cite{shen2017boundary}, researchers proposed a boundary aware fully convolutional neural network, including two branches for up-sampling. The boundary detection branch aims to learn and model boundary information of the whole tumor as a binary classification problem. The region detection branch learns to detect and classify sub-region classes of the tumor. The outputs from the two branches are concatenated and fed to a block of two convolutional layers with a softmax classification layer.

One important mutant of FCN is U-Net \cite{ronneberger2015u}. U-Net consists of a contracting path to capture features and a symmetric expanding path that enables precise localisation. One advantage of using U-Net, compared against traditional FCN, is the skip connections between the contracting and the expanding paths. The skip connections pass feature maps from the contracting path to the expanding path and concatenate the feature maps from the two paths directly. The original image data through skip connections can help the layers in the contracting path recover details. Several research works have been proposed for brain tumor segmentation based on U-Net. For example, Brosch et al. \cite{brosch2016deep} used a fully convolutional network with skip connections to segment multiple sclerosis lesions. Isensee et al. \cite{isensee2017brain} reported a modified U-Net for brain tumor segmentation, where the authors used a dice loss function and extensive data augmentation to successfully avoid over-fitting. In \cite{dong2017automatic}, the authors used zero padding to keep the identical output dimension for all the convolutional layers in both down-sampling and up-sampling paths. Chang et al. \cite{chang2016fully} reported a fully convolutional neural network with residual connections. Similar to skip connection, the residual connection allows both low- and high-level feature maps to contribute towards the final segmentation.

In order to extract information from the original volumetric data, Milletari et al. \cite{milletari2016v} introduced a modified 3D version of U-Net, called V-Net, with a customized dice coefficient loss function. Beers et al. \cite{beers2017sequential} introduced 3D U-Nets based on sequential tasks, which uses the entire tumor ground truth as an auxiliary channel to detect enhancing tumors and tumor cores. In the post-processing stage, the authors employed two additional U-Nets that serve to enhance prediction for better classification outcomes. The input patches consist of seven channels: four anatomical MR and three label maps corresponding to the entire tumor, enhancing tumor, and tumor core.

\setlength{\rotFPtop}{0pt plus 1fil}
\begin{sidewaystable*}
\caption{Comparison between novel methods focuses on effective network design. We categorise the methods based on their main contributions. In column \textbf{Input}, 'P' means patch and 'I' means image. '\textbf{Dim}' means the dimension of the network. In column \textbf{Loss}, 'CE' means cross-entropy loss, 'mIoU' means the mean Intersection of Union and 'KL' means KL-divergence. In column \textbf{Dice} and \textbf{Hausdorff}, 'WT' means whole tumor, 'TC' means tumor core and 'ET' means enhancing tumor. Column \textbf{Dataset} indicates the associated dataset with the segmentation performance. In column Type, 'CV' means cross-validation on the BraTS training set, 'V' means BraTS validation set and 'T' means BraTS test set. '-' means the entry has not been reported in the original paper.}
\label{table:effectivenetwork}

\centering
\begin{tabular}{|>{\centering}p{0.18\linewidth}|p{0.03\linewidth}|c|c|c|c|c|c|c|c|c|c|c|} 
\hline
\multicolumn{2}{|c|}{\multirow{2}{*}{\textbf{Methods}}}         & \multirow{2}{*}{\textbf{Input}} & \multirow{2}{*}{\textbf{Dim}} & \multirow{2}{*}{\textbf{Loss}} & \multicolumn{3}{c|}{\textbf{Dice}}                  & \multicolumn{3}{c|}{\textbf{Hausdorff}}             & \multicolumn{2}{c|}{\textbf{Dataset}}  \\ 
\cline{6-13}
\multicolumn{2}{|c|}{}                                 &                        &                            &                       & WT & TC & ET & WT & TC & ET & Year & Type                   \\ 
\hline
\multirow{14}{*}{\shortstack{Designing\\Effective\\Modules}}&

\cite{dvovrak2015local} & P                      & 2D                         & -                     & 0.81        & 0.79       & -               & -           & -          & -               & 2014 & V                      \\ 
\cline{2-13}
                                                    & \cite{zhao2018deep} & I                      & 2D                         & Softmax               & 0.84        & 0.73       & 0.62            & -           & -          & -               & 2015 & T                      \\ 
\cline{2-13}
                                                    & \cite{rao2015brain} & P                      & 2D                         & -                     & -           & -          & -               & -           & -          & -               & -    & -                      \\ 
\cline{2-13}
                                                    & \cite{casamitjana20163d} & I                      & 3D                         & CE                    & 0.91        & 0.83       & -               & -           & -          & -               & 2015 & CV                     \\ 
\cline{2-13}
                                                    & \cite{shaikh2017brain} & I                      & 2D                         & CE+Soft Dice          & 0.87        & 0.68       & -               & -           & -          & -               & 2017 & V                      \\ 
\cline{2-13}
                                                    & \cite{ghaffari2020brain} & I                      & 3D                         & Dice                  & 0.9         & 0.82       & 0.78            & 5.14        & 6.64       & 7.71            & 2020 & V                      \\ 
\cline{2-13}
                                                    & \cite{zhou2020afpnet} & I                      & 2D                         & -                     & 0.86        & 0.77       & 0.74            & -           & -          & -               & 2018 & V                      \\ 
\cline{2-13}
                                                    & \cite{wang2020modality} & I                      & 2D                         & CE+Dice+MP            & 0.91        & 0.85       & 0.79            & 4.71        & 5.7        & 35.01           & 2020 & V                      \\ 
\cline{2-13}
                                                    & \cite{andermatt2016multi} & P                      & 3D                         & Multinomial Logistic  & -           & -          & -               & -           & -          & -               & -    & -                      \\ 
\cline{2-13}
                                                    & \cite{cheng2019memory} & I                      & 2D                         & Dice+Edge+Mask        & 0.9         & 0.82       & 0.78            & 5.41        & 7.26       & 5.282           & 2019 & V                      \\ 
\cline{2-13}
                                                    & \cite{brugger2019partially}& I                      & 3D                         & Dice                  & 0.91        & 0.86       & 0.81            & 5.61        & 7.83       & 3.35            & 2018 & V                      \\ 
\cline{2-13}
                                                    & \cite{pendse2020memory} & I                      & 3D                         & -                     & -           & -          & -               & -           & -          & -               & -    & -                      \\ 
\cline{2-13}
                                                    & \cite{nuechterlein20183d} & I                      & 2D                         & mIoU                  & 0.85        & 0.78       & 0.67            & 9.6         & 8.67       & 5.5             & 2018 & T                      \\ 
\cline{2-13}
                                                    & \cite{chen20193d} & I                      & 3D                         & Generalized Dice      & 0.91        & 0.85       & 0.8             & 4.66        & 6.44       & 3.06            & 2018 & V                      \\ 
\hline
\multirow{22}{*}{\shortstack{Designing\\Effective\\Architectures}}& \cite{zikic2014segmentation} & P                      & 2D                         & Log Loss              & 0.84        & 0.73       & 0.69            & -           & -          & -               & 2013 & V                      \\ 
\cline{2-13}
                                                    & \cite{pereira2016brain} & P                      & 2D                         & Categorical CE        & 0.78        & 0.65       & 0.75            & 15.83       & 26.54      & 6.99            & 2015 & T                      \\ 
\cline{2-13}
                                                    & \cite{havaei2017brain} & P                      & 2D                         & Surrogate~Loss        & 0.84        & 0.71       & 0.57            & -           & -          & -               & 2013 & T                      \\ 
\cline{2-13}
                                                    & \cite{castillo2017volumetric} & P                      & 3D                         & -                     & -           & -          & -               & -           & -          & -               & 2015 & CV                     \\ 
\cline{2-13}
                                                    & \cite{kamnitsas2017ensembles} & P                      & 3D                         & CE/IoU                & 0.9         & 0.8        & 0.74            & 4.23        & 6.56       & 4.5             & 2017 & V                      \\ 
\cline{2-13}
                                                    & \cite{zhao2016multiscale} & P                      & 2D                         & -                     & -           & -          & -               & -           & -          & -               & -    & -                      \\ 
\cline{2-13}
                                                    & \cite{jesson2017brain} & I                      & 3D                         & Categorical CE        & 0.9         & 0.75       & 0.71            & 4.16        & 8.65       & 6.98            & 2017 & V                      \\ 
\cline{2-13}
                                                    & \cite{isensee2017brain} & I                      & 3D                         & Dice                  & 0.9         & 0.8        & 0.73            & 7           & 9.48       & 4.55            & 2017 & V                      \\ 
\cline{2-13}
                                                    & \cite{dong2017automatic} & I                      & 2D                         & Soft Dice             & 0.86        & 0.86       & 0.65            & -           & -          & -               & 2015 & CV                     \\ 
\cline{2-13}
                                                    & \cite{chang2016fully} & I                      & 2D                         & -                     & 0.89        & 0.83       & 0.78            & 8           & 10         & 5.9             & 2016 & T                      \\ 
\cline{2-13}
                                                    & \cite{chen2017brain} & P                      & 2D                         & KL                    & 0.87        & 0.74       & 0.65            & -           & -          & -               & 2017 & V                      \\ 
\cline{2-13}
                                                    & \cite{chen2019dual} & P                      & 3D                         & KL                    & 0.89        & 0.74       & 0.73            & -           & -          & -               & 2017 & V                      \\ 
\cline{2-13}
                                                    & \cite{pawar2017residual} & I                      & 2D                         & Softmax               & 0.82        & 0.63       & 0.57            & -           & -          & -               & 2017 & V                      \\ 
\cline{2-13}
                                                    & \cite{chen2018s3d} & I                      & 3D                         & Dice                  & 0.84        & 0.78       & 0.68            & 9.2         & 7.71       & 4.52            & 2018 & T                      \\ 
\cline{2-13}
                                                    & \cite{fang2018three} & I                      & 2D                         & -                     & 0.86        & 0.73       & 0.72            & 7.5         & 9.5        & 5.7             & 2018 & V                      \\ 
\cline{2-13}
                                                    & \cite{hua2018multimodal} & I                      & 3D                         & Focal                 & 0.9         & 0.84       & 0.77            & 5.18        & 6.28       & 3.51            & 2018 & V                      \\ 
\cline{2-13}
                                                    & \cite{isensee2018no} & P                      & 3D                         & Dice                  & 0.91        & 0.86       & 0.81            & 4.27        & 6.52       & 2.41            & 2018 & V                      \\ 
\cline{2-13}
                                                    & \cite{li2018fused} & I                      & 3D                         & Dice                  & 0.88        & 0.79       & 0.72            & 29.21       & 11.06      & 7.93            & 2018 & V                      \\ 
\cline{2-13}
                                                    & \cite{myronenko20183d} & I                      & 3D                         & Dice+L2+KL            & 0.91        & 0.87       & 0.82            & 4.52        & 6.85       & 3.92            & 2018 & V                      \\ 
\cline{2-13}
                                                    & \cite{zhao2019bag} & P                      & 3D                         & CE+Dice               & 0.91        & 0.84       & 0.75            & 4.57        & 5.58       & 3.84            & 2019 & V                      \\ 
\cline{2-13}
                                                    & \cite{Yuan2020AutomaticBT} & I                      & 3D                         & Jaccard+Focal         & 0.91        & 0.85       & 0.79            & 4.09        & 5.88       & 18.19           & 2020 & V                      \\ 
\cline{2-13}
                                                    & \cite{henry2020brain} & I                      & 2D                         & Dice                  & 0.91        & 0.85       & 0.8             & 4.3         & 5.69       & 20.56           & 2020 & V                      \\
\hline
\end{tabular}
\end{sidewaystable*}

\subsection{Summary}
In this section, we review and compare the work focused on module and network architecture design in brain tumor segmentation. Table \ref{table:effectivenetwork} shows the results generated by methods focused on module and network architecture design in brain tumor segmentation. We drawn key information of these research works and list it below.
\begin{enumerate}

    \item By designing custom modules, the accuracy and speed of the network can be improved.
    
    \item By designing a customised architecture, it can help the network learn features at different scales, which is one of the most important steps to achieve accurate brain tumor segmentation.

    \item The design of modules and networks heavily relies on human experience. In the future, we anticipate the application of network architecture search for searching effective brain tumor segmentation architectures \cite{bae2019resource}, \cite{kim2019scalable}, \cite{zhou2019unet++}, \cite{zhu2019v}.

    \item Most of the existing network architecture designs do not combine domain knowledge about brain tumor, such as modelling degree information and physically inspired morphological information within tumor segmentation network.
\end{enumerate}

\section{Segmentation under Imbalanced Condition}
\label{sec:Segmentation under Imbalanced Condition}

One of the long standing challenges for brain tumor segmentation is the data imbalance issue. As shown in Fig \ref{fig:challenge} (c), imbalance is mainly reflected in the number of pixels in the sub-regions of the brain tumor. In addition, there is also an imbalance issue in patient samples, that is, the number of the HGG cases is much more than that of the LGG cases. At the same time, labeling biases introduced by manual experts can also be treated as a special form of data imbalance (different experts have different standards, resulting in imbalanced labeling results). Data imbalance plays a significant effect on learning algorithms especially deep networks. The main manifestation is that learning models trained with imbalanced data tend to learn more about the dominant groups, e.g. to learn the morphology of the edema area, and to learn HGG instead of LGG patients) \cite{dong2018imbalanced}, \cite{johnson2019survey}, \cite{sudre2017generalised}.

It is less likely to deal with the data imbalance issue by designing a specific module or architecture. Numerous works have presented many improvement strategies to address data imbalance. According to core components of these strategies, we divide the existing methods into three categories: \textbf{multi-network driven}, \textbf{multi-task driven} and \textbf{custom loss function driven} approaches.

\subsection{Multi-Network Driven Approaches}
Even if complex modules and architectures have been designed to ensure the learning of high-dimensional discriminative features, a single network often suffers from the problem of data imbalance. Inspired by the methods such as multi-expert systems, people have started to construct complex network systems to effectively deal with data imbalance and achieved promising segmentation performance. Common multi-network systems can be divided into network cascade and network ensemble, according to data flows shared between multiple networks.

\begin{figure}[t]
    \centering
    \includegraphics[width=0.5\textwidth]{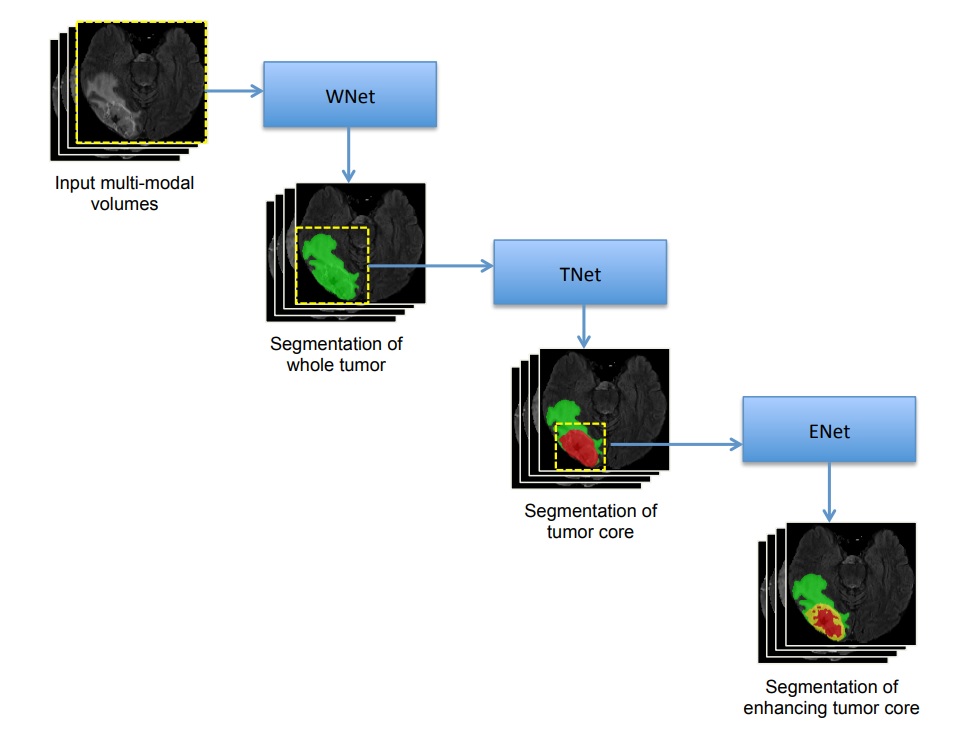}
    \caption{The structure of cascaded convolutional networks for brain tumor segmentation, modified from the original structure reported in \cite{wang2017automatic}. WNet, TNet and ENet are used for segmenting the whole tumor, tumor core and enhancing tumor core, respectively.}
    \label{fig:cascaded}
\end{figure}

\subsubsection{Network Cascade}
The definition of network cascade is that, in a serially connected network, the output of an upstream network is passed to the downstream network as input. This topology simulates the coarse-to-fine strategy, that is, the upstream network extracts rough information or features, and the downstream network subdivides the input and achieves a fine-grained segmentation. 

The earliest work of adopting the cascade strategy was undertaken by Wang et al \cite{wang2017automatic} (Fig. \ref{fig:cascaded}). In their work, the author proposed to connect three networks in series. First, WNet segmented Whole Tumor, and output the segmentation result of Whole Tumor to TNet, and TNet traces Tumor Core. Finally, the segmentation result of TNet is handed over to ENet for the segmentation of Enhancing Tumor. This design logic is inspired by the attributes of the tumor sub-region, where it is assumed that Whole Tumor, Tumor Core, and Enhancing Tumor are included one by one. Therefore, the segmentation output of the upstream network is the Region-of-Interest (RoI) of the downstream network. The advantage of this practice is to avoid the interference caused by the unbalanced data. The introduction of astropic convolution and the manually cropped input effectively reduces the amount of network parameters. But there are two disadvantages: First of all, the segmentation effect of the downstream network is heavily dependent on the performance of the upstream network. Second, only the upstream segmentation result is considered as the input so that the downstream network cannot use other image areas as auxiliary information, which is not conducive to other tasks such as tumor location detection. Similarly, Hua et al. \cite{hua2018multimodal} also proposed a network cascade based on the physical inclusion characteristics of tumor. Unlike Wang et al. \cite{wang2017automatic}, \cite{hua2018multimodal} replaced the cascade unit with a V-Net, which is suitable for 3D segmentation to improve performance. Fang et al. \cite{fang2018three} trained two networks to act as upstream networks at the same time according to different organisational characteristics highlighted by different modalities, respectively training for Flair and T1ce, and then the results of the two upstream networks can be passed to the downstream network for final fine segmentation. Jia et al. \cite{jia2020h2nf} replaced upstream and downstream networks with HRNet \cite{sun2019deep} to preserve maximum spatial resolutions.

Combining 3D networks for cascading can bring better segmentation performance, but the combination of multiple 3D networks requires a large amount of parameters and high computational resources. In response to this, Li \cite{li2018fused} proposed a cascading model that mixes 2D and 3D networks. 2D networks learn from multi views of a volume to obtain the segmentation mask of the whole tumor. Then, the whole tumor mask and the original 3D volume are fed to the downstream 3D U-Net. The downstream network pairs tumor core and enhancing tumor for fine segmentation. Li et al. \cite{li2019multi} also adopted a similar method by connecting multiple U-Nets in series for coarse-to-fine segmentation. The segmentation results at each stage is associated with different loss functions. Vu et al. \cite{vu2019tunet} further introduced dense connection between the upstream and downstream networks to enhance feature expression. The two-stage cascaded U-Net designed by Jiang et al. \cite{jiang2019two} has been further enhanced at the output end. In addition to the single network architecture, they also tried two different segmentation modules (interpolation and deconvolution) at the output end.

In addition to coarse-to-fine segmentation, there are other attempts to introduce other auxiliary functions. Liu designed a novel strategy in \cite{liu2020automatic} to pass the segmentation result of the upstream network to the downstream network. The downstream network reconstructs the original input image according to the segmentation result of the upstream network. The loss of the recovery network is also back-propagated to the upstream segmentation network, in order to help the upstream network to outline the tumor area. Cirillo et al. \cite{cirillo2020vox2vox} introduced adversarial training to tumor segmentation. The generator network constitutes the upstream network, and the discriminator network is used as the downstream network to determine whether a segmentation map is from ground truth or not. Chen et al. \cite{chen2020brain} introduced left and right symmetry characteristics of the brain to the system. The added left and right similarity masks at the connection of the upstream and downstream networks can improve the robustness of network segmentation.

\subsubsection{Network Ensemble}
One main drawback of using a single deep neural network is that its performance is heavily influenced by the hyper-parameter choices. This refers to a limited generalisation capability of the deep neural network. Cascaded network intends to aggregate multiple networks' output in a coarse-to-fine strategy, however downstream networks' performance relies on the upstream network, which still limits the capability of a cascaded system. In order to achieve a more robust and more generalised tumor segmentation, the segmentation output from multiple networks can be aggregated together with a high variance, known as network ensemble. Network ensemble enlarges the hypothesis space of the parameters to be trained by aggregating multiple networks and avoids falling into local optimum caused by data imbalance.

Early research works presented in multi-path network (Sec. \ref{subsubsection:Multi-Path Architecture}) such as Castillo et al. \cite{castillo2017volumetric}, Kamnitsas et al. \cite{kamnitsas2016deepmedic}, \cite{kamnitsas2017efficient} can be regarded as a special form of network ensemble, where each path can be treated as a sub-network. The features extracted by the sub-network are then ensembled and processed for the final segmentation. In this section, we pay more attention to explicit ensemble of segmentation results from multiple sub-networks, rather than implicit ensemble of the features extracted by sub-paths.

Ensembles of multiple models and architectures (EMMA) \cite{kamnitsas2017ensembles} is one of the early well-structured works using ensemble deep neural networks for brain tumor segmentation. EMMA ensembles segmentation results from DeepMedic \cite{kamnitsas2016deepmedic}, FCN \cite{long2015fully} and U-Net \cite{ronneberger2015u} and associated the final segmentation with the highest confidence score. Kao et al. \cite{kao2018brain} ensembles 26 neural networks for tumor segmentation and survival prediction. \cite{kao2018brain} introduced brain parcellation atlas to produce a location prior information for tumor segmentation. Lachinov et al. \cite{lachinov2019knowledge} ensembles two variant U-Net \cite{isensee2018no}, \cite{myronenko20183d} and a cascaded U-Net \cite{lachinov2018glioma}. The final ensemble result out-performs each single network $1-2\%$. 

\begin{figure*}[t]
    \centering
    \includegraphics[width=0.8\textwidth]{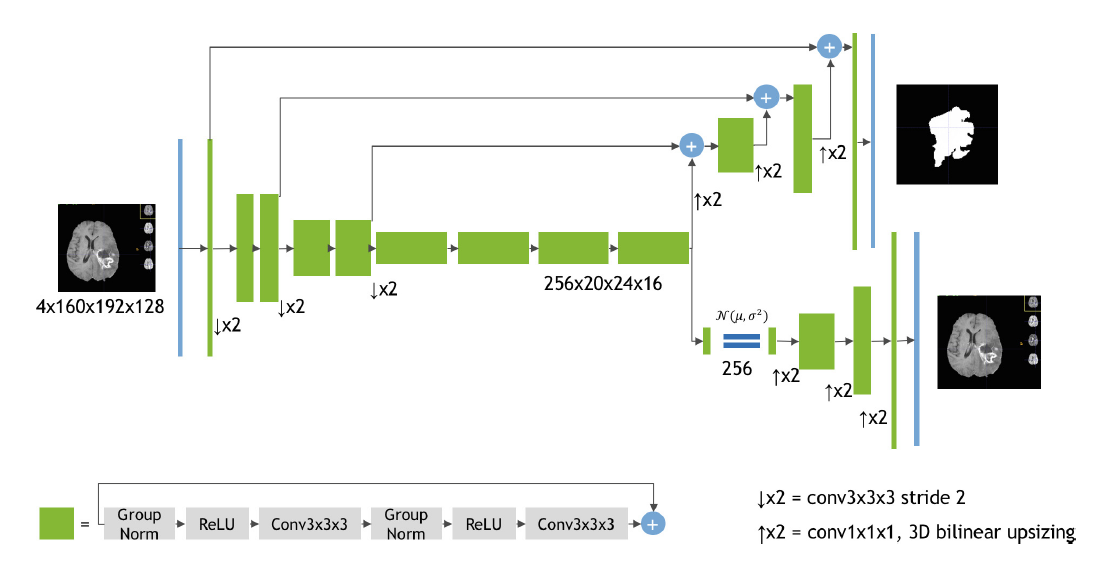}
    \caption{The structure of multi-task networks for brain tumor segmentation. Image courtesy of \cite{myronenko20183d}. The shared encoder learns generalised feature representation and the reconstruction decoder performs multi-task as regularisation.}
    \label{fig:multitask}
\end{figure*}

Instead of feeding sub-networks with the same input, Zhao et al. \cite{zhao20173d} averaged ensembles 3 2D-FCNs where each FCN takes different view slices as input. Similarly, Sundaresan et al. \cite{sundaresan2020brain} averaged ensembles 4 2D-FCNs, where each FCN is designed for segmenting a specific tumor region. Chen et al. \cite{chen2017fully} used a DeconvNet \cite{noh2015learning} to generate a primary segmentation probability map and another multi-scale Convolutional Label Evaluation Net is used to evaluate previously generated segmentation maps. False positives can be reduced using both the probability map and the original input image. Hu et al. \cite{hu20173d} ensembles a 3D cascaded U-Net with a multi-modality fusion structure. The proposed two-level U-Net in \cite{hu20173d} aims to outline the boundary of tumors and the patch-based deep network associates tumor voxels with predicted labels.

Ensemble can be regarded as a boosting strategy for improving final segmentation results by aggregating results from multiple homogeneous networks. The winner of the BraTS2018 \cite{myronenko20183d} ensembles 10 models, which further boosted the performance with $1\%$ on dice score compared with the best single network segmentation. Similar benefits brought by ensembling can be observed from Silva et al. \cite{silva2020multi} as well. BraTS2019 winner \cite{jiang2019two} also adopted an ensemble strategy where the final result is generated by ensembling 12 models, which slightly improves the result (around $0.6-1\%$) compared with the best single model's performance.

\subsection{Multi-Task Driven Approaches}
Most of the work described above only perform single-task learning, that is, only design and optimise a network for precise segmentation of brain tumors only. The disadvantage of single-task learning is that the training target of a single-task may ignore the potential information in some tasks. Information from related tasks may improve the performance of tumor segmentation. Therefore, in recent years, many research works have started from the perspective of multi-task learning, introducing auxiliary tasks on the basis of precise segmentation of brain tumors. The main setting of multi-task learning is a low-level feature representation that can be shared among multiple tasks. There are two advantages from the shared representation. One is to share the learnt domain-related information with each other through shallow shared representations so as to promote learning and to enhance the ability to obtain updated information. The second is mutual restraint. When multi-task learning performs gradient back-propagation, it will take into account the feedback of multiple tasks. Since different tasks may have different noise patterns, the model that learns multiple tasks at the same time will learn a more general representation, which reduces the risk of overfitting and increases the generalisation ability of the system. 

Early attempts such as \cite{zhou2018learning} and \cite{shen2017multi} adapt the idea of multi-task learning and split the brain tumor segmentation task into three different sub-region segmentation tasks, i.e. segmenting whole tumor, tumor core and enhancing tumor individually. In \cite{zhou2018learning}, the author incorporated three sub-region segmentation tasks into an end-to-end holistic network, and exploited the underlying relevance among the three sub-region segmentation tasks. In \cite{shen2017multi}, the author designed three different loss functions, corresponding to the segmentation loss of whole tumor, tumor core and enhancing tumor. In addition, more recent works introduce auxiliary tasks different from image segmentation. The learnt features from other tasks will support accurate segmentation. In \cite{shen2017boundary}, the author additionally introduces a boundary localisation task. The features extracted by the shared encoder are not only suitable for tumor segmentation, but also for tumor boundary localisation. Precise boundary localisation can assist in minimising the searching space and defining precise boundaries during tumor segmentation. \cite{nguyen2020enhancing} introduced the idea of first detecting and then segmenting, that is, detecting the location of tumors, and then performing precise tumor segmentation.

Another commonly used auxiliary task is to reconstruct the input data, that is, the learnt feature representation can be restored to the original input by an auxiliary decoder. \cite{myronenko20183d} is the first method to introduce reconstruction as an auxiliary task to brain tumor segmentation. \cite{weninger2019multi} introduced two auxiliary tasks, reconstruction and enhancement, to further enhance the ability of feature representation. \cite{iwasawa2020label} introduced three auxiliary tasks, including reconstruction, edge segmentation and patch comparison. These works regard the auxiliary task as a regularization to the main brain tumor segmentation task. Most multi-task designs use shared encoder to extract features and independent decoders to process different tasks. From the perspective of parameter update, the role of auxiliary task is to further regularize shared encoder's parameter. Different from L1 or L2 that explicitly regularize parameter numbers and values, the auxiliary task shared low-level sub-space with main task. During training, auxiliary task is helpful for the network to train in the direction that simultaneously optimize the auxiliary task and the main segmentation task, which reduces the search space of the parameters, makes the extracted features more generalized for accurate segmentation \cite{caruana1997multitask}, \cite{evgeniou2004regularized}, \cite{sener2018multi}, \cite{zhang2021survey}.

\subsection{Customised Loss Function Driven Approaches}

During network training, the gradient is likely dominated by the excessively large sample if we use an imbalanced dataset. Therefore, a number of works propose a custom loss function to regulate gradients during the training of a brain tumor segmentation model. Designing a custom loss function aims to reduce the weights of the easy-to-classify samples in the loss function, whilst increasing the weights of the hard samples, so that the model is more focused on the samples of a small proportion, reducing the impact of imbalanced datasets.

\setlength{\rotFPtop}{0pt plus 1fil}
\begin{sidewaystable*}
\caption{Comparisons between the methods with the novelty of dealing with the imbalance issue. We categorise each method based on its main novel contribution. In column \textbf{Input}, 'P' means the patch and 'I' means the image. '\textbf{Dim}' means the dimension of the network. \textbf{\#Nets} means the number of the network candidates. In column \textbf{Connection}, 'C' means the cascade connection and 'E' means the network ensemble. In column \textbf{Task}, 'S' means the segmentation task, 'G' means the modality generation task, 'C' means the classification task, 'B' means the boundary segmentation task, 'R' means the input reconstruction task. In column \textbf{Loss}, 'CE' means the cross-entropy loss, 'TV' means the total variation loss and 'KL' means the KL-divergence, 'CPC' means the contrastive predictive coding loss. In column \textbf{Dice} and \textbf{Hausdorff}, 'WT' means whole tumor, 'TC' means tumor core and 'ET' means enhancing tumor. Column \textbf{Dataset} indicates the associated dataset with the reported segmentation performance. In column Type, 'CV' means the cross-validation on BraTS training set, 'V' means the BraTS validation set and 'T' means the BraTS test set. '-' means the entry was not reported in the original paper.}
\label{table:imbalancesegmentation}
\centering
\begin{tabular}{|>{\centering}p{0.08\linewidth}|p{0.02\linewidth}|c|c|c|c|c|c|c|c|c|c|c|c|c|c|} 
\hline
\multicolumn{2}{|c|}{\multirow{2}{*}{\textbf{Methods}}}                   & \multirow{2}{*}{\textbf{Input}} & \multirow{2}{*}{\textbf{Dim}} & \multirow{2}{*}{\textbf{\#Nets}} & \multirow{2}{*}{\textbf{Connection}} & \multirow{2}{*}{\textbf{Task}} & \multirow{2}{*}{\textbf{Loss}} & \multicolumn{3}{c|}{\textbf{Dice}}                  & \multicolumn{3}{c|}{\textbf{Hausdorff}}             & \multicolumn{2}{c|}{\textbf{Dataset}}  \\ 
\cline{9-16}
\multicolumn{2}{|c|}{}                                           &                        &                            &                             &                             &                       &                       & WT & TC & ET & WT & TC & ET & Year & Type                   \\ 
\hline
\multirow{15}{*}{\shortstack{Multi\\Networks\\Driven\\Approaches}}& \cite{wang2017automatic} & P                      & 3D                         & 3                           & C                           & S                     & Dice                  & 0.9         & 0.84       & 0.78            & 3.89        & 6.48       & 3.28            & 2017 & V                      \\ 
\cline{2-16}
                                                              & \cite{hua2018multimodal} & P                      & 3D                         & 5                           & C+E                          & S                     & Focal                 & 0.9         & 0.84       & 0.77            & 5.18        & 6.28       & 3.51            & 2018 & V                      \\ 
\cline{2-16}
                                                              & \cite{fang2018three} & I                      & 2D                         & 2                           & E                           & S                     & -                     & 0.86        & 0.73       & 0.72            & 7.5         & 9.5        & 5.7             & 2018 & V                      \\ 
\cline{2-16}
                                                              & \cite{jia2020h2nf} & I                      & 3D                         & 2                           & C                           & S                     & Dice + CE             & 0.91        & 0.85       & 0.79            & 4.18        & 4.97       & 26.57           & 2020 & V                      \\ 
\cline{2-16}
                                                              & \cite{li2018fused} & I                      & 3D                         & 5                           & C+E                          & S                     & Dice                  & 0.88        & 0.79       & 0.72            & 29.21       & 11.06      & 7.93            & 2018 & V                      \\ 
\cline{2-16}
                                                              & \cite{vu2019tunet} & I                      & 3D                         & 3                           & C                           & S                     & Dice                  & 0.9         & 0.81       & 0.78            & 4.32        & 6.28       & 3.7             & 2019 & V                      \\ 
\cline{2-16}
                                                              & \cite{jiang2019two} & I                      & 3D                         & 2                           & C                           & S                     & Dice                  & 0.91        & 0.86       & 0.8             & 4.26        & 5.43       & 3.14            & 2019 & V                      \\ 
\cline{2-16}
                                                              & \cite{liu2020automatic} & I                      & 2D                         & 2                           & C                           & G+S                   & L1+TV                 & -           & -          & -               & -           & -          & -               & -    & -                      \\ 
\cline{2-16}
                                                              & \cite{cirillo2020vox2vox} & I                      & 3D                         & 2                           & C                           & G+S                   & L2+Dice               & 0.89        & 0.79       & 0.75            & 6.39        & 14.07      & 36              & 2020 & V                      \\ 
\cline{2-16}
                                                              & \cite{kamnitsas2017ensembles} & I                      & 3D                         & 3                           & E                           & S                     & CE+IoU                & 0.9         & 0.8        & 0.74            & 4.23        & 6.56       & 4.5             & 2017 & V                      \\ 
\cline{2-16}
                                                              & \cite{lachinov2019knowledge} & I                      & 3D                         & 3                           & E                           & S                     & Dice+CE               & 0.9         & 0.84       & 0.76            & -           & -          & -               & 2019 & V                      \\ 
\cline{2-16}
                                                              & \cite{zhao20173d} & P                      & 2D                         & 3                           & E                           & S                     & -                     & 0.89        & 0.79       & 0.75            & -           & -          & -               & 2017 & V                      \\ 
\cline{2-16}
                                                              & \cite{sundaresan2020brain} & P                      & 2D                         & 4                           & E                           & S                     & Dice+CE               & 0.89        & 0.77       & 0.77            & 4.4         & 15.3       & 29.4            & 2020 & V                      \\ 
\cline{2-16}
                                                              & \cite{hu20173d} & I                      & 2D                         & 3                           & C                           & S+C                   & Dice                  & 0.85        & 0.7        & 0.65            & 25.24       & 21.45      & 17.98           & 2017 & V                      \\ 
\cline{2-16}
                                                              & \cite{silva2020multi} & P                      & 2D                         & 3                           & C                           & S                     & CE                    & 0.91        & 0.81       & 0.76            & 4.34        & 9.39       & 27.16           & 2020 & V                      \\ 
\hline
\multirow{7}{*}{\shortstack{Multi\\Tasks\\Driven\\Approaches}}& \cite{zhou2018learning} & P                      & 3D                         & 2                           & C+E                          & S+C                   & -~                    & 0.91        & 0.86       & 0.81            & 4.17        & 6.54       & 2.71            & 2018 & V                      \\ 
\cline{2-16}
                                                              & \cite{chen2020brain} & -                      & -                          & 1                           & -                           & S+Sim                 & CE+Focal              & 0.85        & 0.68       & 0.58            & -           & -          & -               & 2015 & V                      \\ 
\cline{2-16}
                                                              & \cite{shen2017multi} & I                      & 2D                         & 1                           & -                           & S                     & CE                    & 0.88        & 0.71       & 0.73            & -           & -          & -               & 2015 & T                      \\ 
\cline{2-16}
                                                              & \cite{shen2017boundary} & I                      & 2D                         & 1                           & -                           & S+B                   & CE                    & 0.89        & 0.72       & 0.73            & -           & -          & -               & 2015 & T                      \\ 
\cline{2-16}
                                                              & \cite{myronenko20183d} & I                      & 3D                         & 10                          & E                           & S+R                   & Dice+L2+KL            & 0.91        & 0.87       & 0.82            & 4.52        & 6.85       & 3.92            & 2018 & V                      \\ 
\cline{2-16}
                                                              & \cite{weninger2019multi} & P                      & 3D                         & 1                           & -                           & S+R+C                 & Dice+L2+KL            & 0.85        & 0.78       & 0.75            & 7.98        & 8.25       & 5.76            & 2019 & V                      \\ 
\cline{2-16}
                                                              & \cite{iwasawa2020label} & I                      & 3D                         & 1                           & -                           & S+R+B                 & L2+KL+Dice+CE+CPC     & 0.92        & 0.88       & 0.88            & 12.4        & 16.09      & 8.71            & 2017 & Sub                    \\ 
\hline
\multirow{7}{*}{\shortstack{Customized\\Loss\\Function\\Driven\\Approaches}}& \cite{randhawa2016improving} & P                      & 2D                         & 1                           & -                           & S                     & L1+L2+CE              & 0.87        & 0.75       & 0.71            & -           & -          & -               & 2016 & T                      \\ 
\cline{2-16}
                                                              & \cite{liu2020brain} & I                      & 2D                         & 1                           & -                           & S                     & Dice                  & 0.88        & 0.8        & 0.76            & 6.49        & 6.68       & 21.39           & 2020 & V                      \\ 
\cline{2-16}
                                                              & \cite{nguyen2020enhancing} & I                      & 3D                         & 10                          & E                           & S                     & Dice+Focal+CE         & 0.9         & 0.84       & 0.78            & 5.68        & 9.57       & 24.02           & 2020 & V                      \\ 
\cline{2-16}
                                                              & \cite{isensee2017brain} & I                      & 3D                         & 1                           & -                           & S                     & Dice                  & 0.9         & 0.8        & 0.73            & 7           & 9.48       & 4.55            & 2017 & V                      \\ 
\cline{2-16}
                                                              & \cite{jesson2017brain} & I                      & 3D                         & 1                           & -                           & S                     & CE                    & 0.9         & 0.75       & 0.71            & 4.16        & 8.65       & 6.98            & 2017 & V                      \\ 
\cline{2-16}
                                                              & \cite{cheng2019memory} & I                      & 3D                         & 1                           & -                           & S                     & Dice+Edge+Mask        & 0.9         & 0.82       & 0.78            & 5.41        & 7.26       & 5.282           & 2019 & V                      \\ 
\cline{2-16}
                                                              & \cite{pawar2019ensemble} & I                      & 2D                         & 5                           & E                           & S                     & Dice                  & 0.92        & 0.88       & 0.87            & 4.23        & 5.77       & 8.18            & 2019 & V                      \\
\hline
\end{tabular}
\end{sidewaystable*}

Early research works tend to uses the standard loss functions, e.g. categorical cross-entropy \cite{pereira2016brain}, cross-entropy \cite{tseng2017joint}, and dice loss \cite{cata2017masked}. \cite{randhawa2016improving} is the first attempt to customise the loss function. In \cite{randhawa2016improving}, the authors enhance the loss function to give more weights to the edge pixels, which significantly improve segmentation accuracy at classifying tumor boundaries. Experimental results show that the weighted loss function for edge pixels helps to improve the performance of segmentation dice by $2-4\%$. Later on,  \cite{shen2017boundary} proposed a customised cross-entropy loss for boundary pixels while using an auxiliary task that includes boundary localisation. In \cite{liu2020automatic}, the reconstruction task is adopted as regularisation, so the loss function aims at improving pixel-wise reconstruction accuracy. In \cite{liu2020brain}, the space loss function was designed to ensure that the learnt features can keep spatial information as much as possible. \cite{nguyen2020enhancing} further used a focal loss to deal with imbalanced issues. \cite{isensee2017brain} used a multi-class dice loss, that is, the smaller the proportion of the category, the higher the error weight during back-propagation. In \cite{jesson2017brain}, a multi-scale loss function was added to perform in-depth supervision on the features of different scales at each stage of the encoder, helping the network to learn the features in multi-scale resolutions that are more conducive to object segmentation. In \cite{fang2018three}, from the perspective of a modal, two types of losses were designed for T1ce and Flair respectively. \cite{cheng2019memory} proposed a weighted combination of the dice loss, the edge loss and the mask loss. The result shows that the combined losses can improve dice performance by about $2\%$. \cite{pawar2019ensemble} also proposed a combination loss set, which includes the categotical cross-entropy and the soft dice loss.

\begin{figure*}[hbt]
    \centering
    \includegraphics[width=0.8\textwidth]{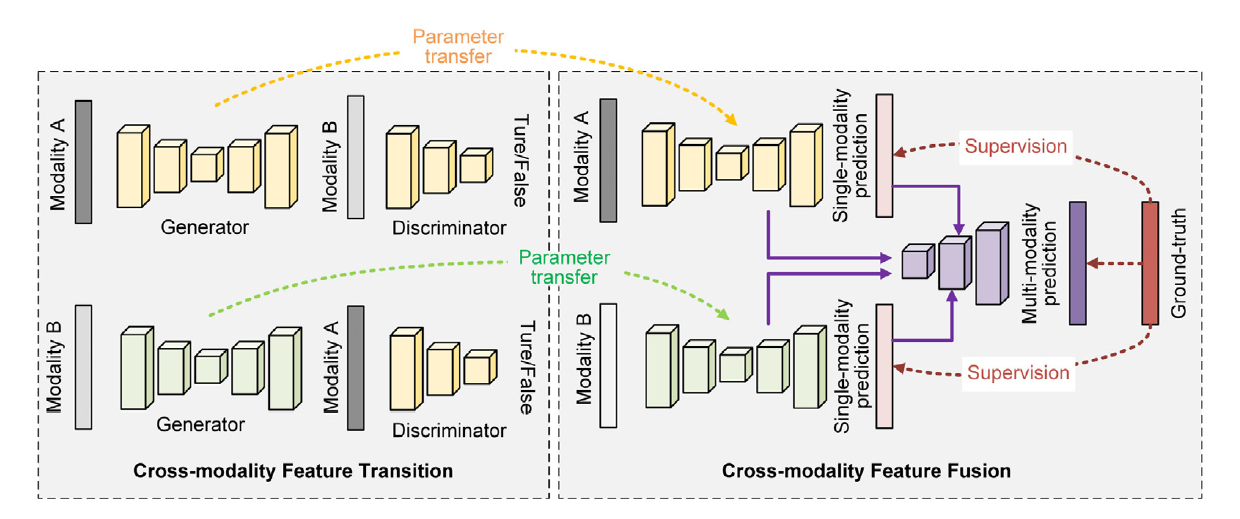}
    \caption{The illustration of cross-modality feature learning framework. Image courtesy from \cite{zhang2020exploring}.}
    \label{fig:modalitypair}
\end{figure*}

\subsubsection{Summary}
Table \ref{table:imbalancesegmentation} shows the results generated by methods focused on dealing with data imbalance in brain tumor segmentation. From the above comparison, we can find several interesting observations.

\begin{enumerate}
    \item From the perspective of the network, the strategy to solve the imbalance problem is mainly to combine the output of multiple networks. Commonly used combination methods include network cascade and network ensemble. But these strategies all depend on the performance of each network. The consumption of the computing resources is also increased proportionally to the number of the network candidates.
    
    \item From the perspective of a task, the strategy to solve the imbalance problem is to set up auxiliary tasks for the regulating networks so that the networks can make full use of the existing data and learn more generalised features that are beneficial to the auxiliary tasks as well as the segmentation task.
    
    \item From the perspective of the loss function, the strategy to solve the imbalance problem is to use a custom loss function or an auxiliary loss function. By weighting the hard samples, the networks are regulated to pay more attention to the small data.
\end{enumerate}

\section{Utilising Multi Modality Information}
\label{sec:Utilizing Multi Modality Information}

Multi-modality imaging has played a key role in medical image analysis and applications. Different modalities of MRI emphasise on different tissues. Effective use of multi-modality information is one of the key factors in MRI-based brain tumor segmentation. According to the number of the available modalities, we divide the multi-modality brain tumor segmentation into two scenes: leveraging information based on multiple modalities and limited information processing with missing modality.

\begin{figure}
    \centering
    \includegraphics[width=0.5\textwidth]{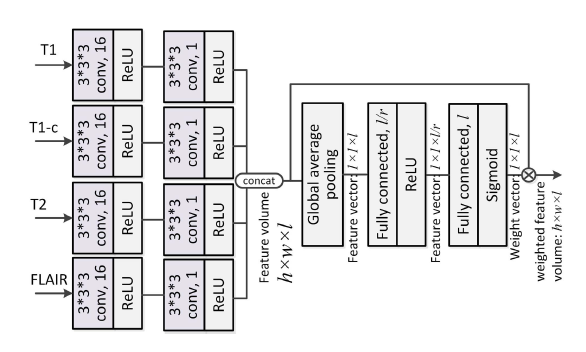}
    \caption{The structure of the modality-aware feature embedding module. Image courtesy of \cite{zhang2021cross}.}
    \label{fig:modalityrank}
\end{figure}

\subsection{Learning with multiple modalities}
In this paper, we follow the BraTS competition standard, that is, multi-modality refers the input data modalities include but not limit to T1, T1ce, T2, and Flair. In order to effectively use multi-modality information, existing works focus on effectively learning multi-modality information. The designed learning methods can be classified into three categories based on their purposes: \textbf{Learning to Rank}, \textbf{Learning to Pair} and \textbf{Learning to Fuse}.

\textbf{Learning to Rank Modalities} 
In multi-modality processing, the existing data modality is sorted by relevance based on the learning task, so that the network can focus on learning the modality with high relevance. This definition can be re-named as modality-task modeling. Early work from \cite{rao2015brain} can be treated as basic learning to rank formation. In \cite{rao2015brain}, the author transformed each modality to a single CNN. In \cite{rao2015brain}, each CNN corresponds to a different modality and the features extracted by CNN are independent of each other. The loss returned by the final classifier is similar to the scoring of the input data and the segmentation is undertaken according to the score. A similar processing method was used in \cite{zhang2021cross}. For each of the two modalities, two independent networks were used for modeling relationship matching, and the parameters of each network are affected by the influence of different supervision losses. \cite{zhang2020exploring} extracted features of different embedding modalities (as shown in Fig. \ref{fig:modalitypair}), modeled the relationship between the modalities and the segmentation of different tumor sub-regions, so that the data of different modalities were weighted and sorted corresponding to individual tasks.

\begin{figure}
    \centering
    \includegraphics[width=0.5\textwidth]{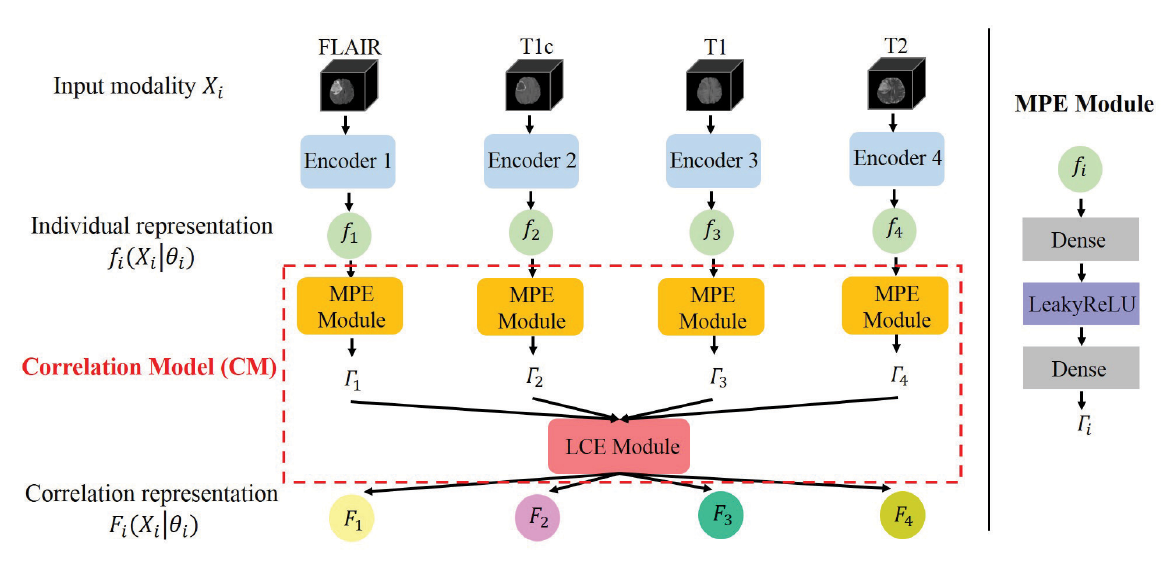}
    \caption{The structure of the modality correlation module. Image courtesy of \cite{zhou2021latent}.}
    \label{fig:missingmodality}
\end{figure}

\textbf{Learning to Pair Modalities}
Learning to rank modalities refers to the sorting of the modality-task relation for a certain segmentation task. Another commonly used modeling is the modality-modality pairing, which selects the best combination from multi-modality data to achieve precise segmentation. \cite{li2017deep} is one of the early works to model the modality-modality relationship. The authors paired every two modalities and sent all the pairing combinations to the downstream network. \cite{zhang2020exploring} further strengthens the modality-modality pairing relationship through the cross-modal feature transition module and the modal pairing module. In the cross-modality feature transition module, the authors converted the input and output from one modality's data to the concatenation of a modality pair. In the cross-modality feature fusion module, the authors converted the single-modality feature learning to the single-modality-pair feature learning, which predicts the segmentation masks of each single-modality-pair.

\textbf{Learning to Fuse Modalities}
More recent works focus on learning to fuse multi-modality. Different from the modality ranking and pairing, modality fusion is to fuse features from each modality for accurate segmentation. The early fusion method is relatively simple, usually concatenates or adds features learned from different modalities. In \cite{rao2015brain}, the authors used 4 networks to extract features from each modality and concatenates the extracted modality aware features. The features after concatenation are sent to Random Forest to classify the central pixel of the input patch. In \cite{fang2018three}, features from T1ce and Flair were added and sent to the downstream network for entire tumor segmentation. Similarly, in \cite{zhang2020exploring}, modality aware feature extraction is performed and sent to the downstream network for further learning. These two fusion methods do not introduce additional parameters and are very simple and efficient. In \cite{zhang2020exploring}, even though the authors fused the features from more complex cross-modal feature pairing and single-modal feature pairing modules. In addition, there are other works such as \cite{tseng2017joint} and \cite{li2017deep} that used additional convolutional modules to combine and learn features from different modalities so as to accomplish modality fusion.

Although concatenation and addition are used, these two fusion methods do not change the semantics of learned features and cannot highlight or suppress features. To tackling this problem, many research works in recent years have adopted attention mechanisms to strengthen the learnt features. \cite{islam2019brain}, \cite{liu2020brain}, \cite{wang2019global} and \cite{zhou2020multi} used a spatial and channel attention based fusion module. The proposed attention mechanism highlights useful features and suppresses redundant features, resulting in accurate segmentation.

\setlength{\rotFPtop}{0pt plus 1fil}
\begin{sidewaystable*}
\caption{Comparison between the methods with the novelty of learning with multi-modality. We categorise each method based on its main novel contribution. In column \textbf{Input}, 'P' means the patch and 'I' means the image. '\textbf{Dim}' means the dimension of the network. Column \textbf{'Learning To'} means the learning task of multi-modality, where  'R' means learning to rank, 'P' means learning to pair, 'F' means learning to fuse, 'G' means learning to generate missing modality, 'Fw/M' means fuse with missing modalities. In column \textbf{Fusion}, 'Concate' means concatenation, 'Conv' means the convolution module, 'Add' mean addition, 'S Att' means spatial attention, 'C Att' means channel attention. In column \textbf{Task}, 'S' means segmentation task, 'G' means modality generation task. In column \textbf{Loss}, 'CE' means cross-entropy loss, 'Adv' means adversarial loss, 'CC' means cycle consistency loss and 'MAE' means mean absolute square loss. In column \textbf{Dice} and \textbf{Hausdorff}, 'WT' means whole tumor, 'TC' means tumor core and 'ET' means enhancing tumor. Column \textbf{Dataset} indicates the associated dataset with the reported segmentation performance. In column Type, 'CV' means cross-validation on the BraTS training set, 'V' means the BraTS validation set and 'Sub' means the manually divided subset from training set. '-' means the entry was not reported in the original paper.}
\label{table:multimodality}
\centering
\begin{tabular}{|>{\centering}p{0.10\linewidth}|p{0.03\linewidth}|c|c|c|c|c|c|c|c|c|c|c|c|c|c|} 
\hline
\multicolumn{2}{|c|}{\multirow{2}{*}{\textbf{Methods}}}         & \multirow{2}{*}{\textbf{Input}} & \multirow{2}{*}{\textbf{Dim}} & \multirow{2}{*}{\textbf{Learning To}} & \multirow{2}{*}{\textbf{Fusion}} & \multirow{2}{*}{\textbf{Task}} & \multirow{2}{*}{\textbf{Loss}} & \multicolumn{3}{c|}{\textbf{Dice}} & \multicolumn{3}{c|}{\textbf{Hausdorff}} & \multicolumn{2}{c|}{\textbf{Dataset}}  \\ 
\cline{9-16}
\multicolumn{2}{|c|}{}                                 &                        &                      &                                               &                                &                       &                       & WT   & TC   & ET          & WT   & TC   & ET               & Year & Type                   \\ 
\hline
\multirow{10}{*}{\shortstack{Learning\\with\\Complete\\Modalities}}& \cite{rao2015brain} & P                      & 2D                   & R+F                                            & Concate                        & S                     & -                     & -    & -    & -           & -    & -    & -                & -    & -                      \\ 
\cline{2-16}
                                                    & \cite{tseng2017joint} & I                      & 2D                   & F                                             & Conv                           & S                     & CE                    & 0.85 & 0.68 & 0.69        & -    & -    & -                & 2015 & V                      \\ 
\cline{2-16}
                                                    & \cite{li2017deep} & I                      & 2D                   & P+F                                            & Conv                           & S                     & Focal                 & 0.88 & 0.71 & 0.75        &      &      &                  & 2017 & V                      \\ 
\cline{2-16}
                                                    & \cite{fang2018three} & I                      & 2D                   & F                                             & Add                       & S                     & -                     & 0.86 & 0.73 & 0.72        & 7.5  & 9.5  & 5.7              & 2018 & V                      \\ 
\cline{2-16}
                                                    & \cite{wang2019global} & I                      & 2D                   & F                                             & S Att + C Att    & S                     & Dice                  & -    & -    & -           & -    & -    & -                & -    & -                      \\ 
\cline{2-16}
                                                    & \cite{islam2019brain} & P                      & 3D                   & F                                             & S Att + C Att    & S                     & -                     & 0.9  & 0.79 & 0.7         & 6.29 & 8.76 & 7.05             & 2019 & V                      \\ 
\cline{2-16}
                                                    & \cite{liu2020brain} & I                      & 2D                   & F                                             & S Att + C Att   & S                     & Dice                  & 0.88 & 0.8  & 0.76        & 6.49 & 6.68 & 21.39            & 2020 & V                      \\ 
\cline{2-16}
                                                    & \cite{zhang2021cross} & P                      & 3D                   & R+P+F                                           & Concate + Add             & S                     & Adv+CC                & 0.9  & 0.84 & 0.79        & 5    & 6.37 & 3.99             & 2018 & V                      \\ 
\cline{2-16}
                                                    & \cite{zhang2020exploring} & P                      & 3D                   & R+F                                            & Concate                        & G+S                    & Dice+T-Test           & 0.9  & 0.82 & 0.78        & 5.73 & 9.27 & 3.57             & 2020 & V                      \\ 
\cline{2-16}
                                                    & \cite{zhou2020multi} & I                      & 3D                   & F                                             & S Att + C Att    & S                     & Dice                  & 0.87 & 0.79 & 0.74        & 7.54 & 7.68 & 6.1              & 2017 & CV                     \\ 
\hline
\multirow{4}{*}{\shortstack{Dealing\\with\\Missing\\Modalities}}& \cite{yu20183d} & I                      & 3D                   & G                                             & -                              & G+S                    & L1                    & 0.68 & 0.72 & -           & -    & -    & -                & 2015 & Sub                    \\ 
\cline{2-16}
                                                    & \cite{zhou2021latent} & I                      & 3D                   & Fw/M                                          & S Att + C Att    & S                     & Dice+MAE              & 0.87 & 0.72 & 0.73        & 6.7  & 9.3  & 6.3              & 2019 & V                      \\ 
\cline{2-16}
                                                    & \cite{zhou2020brain} & I                      & 3D                   & Fw/M                                          & S Att + C Att    & S                     & Dice+MAE              & 0.88 & 0.79 & 0.69        & -    & -    & -                & 2018 & CV                     \\ 
\cline{2-16}
                                                    & \cite{yu2021sa} & I                      & 3D                   & Fw/M                                          & -                              & S                     & -                     & 0.91 & 0.85 & 0.78        & 4.46 & 5.26 & 3.69             & 2019 & V                      \\
\hline
\end{tabular}
\end{sidewaystable*}

\begin{table*}[]
\centering
\caption{Opensourced projects from deep learning based brain tumor segmentation. where '3rd Party' means the code is re-implemented by a third party based on the associated paper.}
\label{table:opensource}
\begin{tabular}{|p{0.46\textwidth}|p{0.46\textwidth}|}
\hline
\multirow{2}{*}{\textbf{Paper   Title}}                                                                                                 & \multirow{2}{*}{\textbf{Code Link}}                                                                                       \\
                                                                                                                               &                                                                                                                  \\ \hline
Brain tumor segmentation with Deep Neural Networks & \begin{tabular}[c]{@{}l@{}}(3rd Party)   https://github.com/naldeborgh7575/ \\brain\_segmentation\end{tabular}

\\ \hline
DeepMedic on   Brain Tumor Segmentation                                                                                        & https://github.com/deepmedic/deepmedic                                                                           \\ \hline
Multi-dimensional   Gated Recurrent Units for Brain Tumor Segmentation                                                         & https://github.com/zubata88/mdgru                                                                                \\ \hline
Volumetric   Multimodality Neural Network For Brain Tumor Segmentation                                                         & https://github.com/BCV-Uniandes/BCVbrats                                                                         \\ \hline
Brain Tumor   Segmentation and Radiomics Survival Prediction: Contribution to the BRATS   2017 Challenge                       & (3rd Party)   https://github.com/pykao/Modified-3D-UNet-Pytorch                                                  \\ \hline
Residual   Encoder and Convolutional Decoder Neural Network for Glioma Segmentation                                            & https://github.com/kamleshpawar17/BratsNet-2017                                                                  \\ \hline
Automatic   Brain Tumor Segmentation Using Cascaded Anisotropic Convolutional Neural   Networks                                & https://github.com/taigw/brats18\_docker                                                                         \\ \hline
No New-Net                                                                                                                     & https://github.com/MIC-DKFZ/nnUNet                                                                               \\ \hline
3D MRI Brain   Tumor Segmentation Using Autoencoder Regularization                                                             & (3rd Party) https://github.com/IAmSuyogJadhav/3d-mri-brain-tumor-segmentation-using-autoencoder-regularization \\ \hline
3D-ESPNet with   Pyramidal Refinement for Volumetric Brain Tumor Image Segmentation                                            & https://github.com/sacmehta/3D-ESPNet                                                                            \\ \hline
One-pass   Multi-task Networks with Cross-task Guided Attention for Brain Tumor   Segmentation                                 & https://github.com/chenhong-zhou/OM-Net                                                                          \\ \hline
Multi-step   Cascaded Networks for Brain Tumor Segmentation                                                                    & https://github.com/JohnleeHIT/Brats2019                                                                          \\ \hline
An Ensemble of   2D Convolutional Neural Network for 3D Brain Tumor Segmentation                                               & https://github.com/kamleshpawar17/Brats19                                                                        \\ \hline
Knowledge   Distillation for Brain Tumor Segmentation                                                                          & https://github.com/lachinov/brats2019                                                                            \\ \hline
Label-Efficient   Multi-Task Segmentation using Contrastive Learning                                                           & https://github.com/pfnet-research/label-efficient-brain-tumor-segmentation                                       \\ \hline
Vox2Vox:   3D-GAN for Brain Tumour Segmentation                                                                                & https://github.com/mdciri/Vox2Vox                                                                                \\ \hline
Brain tumor   segmentation with self-ensembled, deeply-supervised 3D U-net neural networks:   a BraTS 2020 challenge solution. & https://github.com/lescientifik/open\_brats2020                                                                  \\ \hline
Brain tumour   segmentation using a triplanar ensemble of U-Nets on MR images                                                  & https://git.fmrib.ox.ac.uk/vaanathi/truenet\_tumseg                                                              \\ \hline
A Two-Stage   Cascade Model with Variational Autoencoders and Attention Gates for MRI Brain   Tumor Segmentation               & https://github.com/shu-hai/two-stage-VAE-Attention-gate-BraTS2020                                                \\ \hline
HDC-Net:   Hierarchical Decoupled Convolution Network for Brain Tumor Segmentation                                             & https://github.com/luozhengrong/HDC-Net                                                                          \\ \hline
\end{tabular}
\label{tab:opensource}
\end{table*}

\subsection{Dealing with Missing Modalities}
The modality learning methods mentioned above work in the multi-modality scene. For example, in BraTS, we obtain the data of four modalities: T1, T1ce, T2, and FLAIR. However, in actual application scenarios, it is very difficult to obtain complete and high-quality multi-modality datasets, refers to as missing modality scenarios. \cite{yu20183d} is one of the earliest works targeting learning under missing modality. The authors in \cite{yu20183d} constructed the only available modal T1 and used generative adversarial networks to generate the missed modalities. In \cite{yu20183d}, the authors used the existing T1 modality as input to generate Flair modality. The generated Flair data is sent as a supplement with the original T1 data to the downstream segmentation network. \cite{zhou2020brain}, \cite{zhou2021latent} learnt the implicit relationship between modalities and examined all possible missing scenarios. The results show that multi-modality have an important influence on accurate segmentation. In \cite{yu2021sa}, the intensity correction algorithm was proposed for different scenarios of the single modality input. In this framework, the intensity query and correction of the data of multiple modalities makes it easier to distinguish the tumor and non-tumor regions in the synthetic data.

\subsubsection{Summary}

Table \ref{table:multimodality} shows the results generated by methods focused learning with multi-modality in deep learning based brain tumor segmentation. We can collect several common observations in utilising the information from multi modalities.

\begin{enumerate}
    \item For task-modality modeling, learning to rank modalities can help the network choose the most relative and conducive modality for accurate segmentation. Most of the research works model the implicit ranking while learning the modality aware features.
    \item For modality-modality modeling, learning to pair modalities can help the network find the most suitable modality combination for segmentation. However, existing pairing works show modality pairs through exhaustive combination with large computing resources.
    \item The fusion of multi-modality information can improve the expressive ability and generalisation of features. Existing fusion methods have their own advantages and disadvantages. Addition or concatenation does not introduce additional parameters, but lacks the physical expression of features. Using a small network, an attention module can optimise feature expression, but introduce additional parameters and computational cost.
    \item  Missing modalities are one of the common scenes in clinical imaging. Existing works focus on the perspective of generation, using existing modality data to generate missing modalities. However, the performance and quality of the generator modal heavily relies on the quality of the existing modality data.
\end{enumerate}

\section{Conclusion}
\label{sec:Conclusion}
Applying various deep learning methods to brain tumor segmentation is an invaluable and challenging task. Automated image segmentation benefits several aspects due to the powerful feature learning ability of deep learning techniques. In this paper, we have investigated relevant deep learning based brain tumor segmentation methods and presented a comprehensive survey. We structurally categorised and summarised the deep learning based brain tumor segmentation methods. We have widely investigated this task and discussed several key aspects such as methods' pros and cons, designing motivation and performance evaluation.

\section*{Acknowledgement}
This work was funded by the Chine Scholarship Council and Graduate Teaching Assistantship of University of Leicester. Yaochu Jin is supported by an Alexander von Humboldt Professorship endowed by the German Federal Ministry for Education and Research. The authors thank Prof. Guotai Wang, Prof. Dingwen Zhang and Dr. Tongxue Zhou for their detailed suggestions and discussions.

\bibliography{mybibfile}

\begin{thebibliography}{100}
\expandafter\ifx\csname url\endcsname\relax
  \def\url#1{\texttt{#1}}\fi
\expandafter\ifx\csname urlprefix\endcsname\relax\def\urlprefix{URL }\fi
\expandafter\ifx\csname href\endcsname\relax
  \def\href#1#2{#2} \def\path#1{#1}\fi

\bibitem{doi2007computer}
K.~Doi, Computer-aided diagnosis in medical imaging: historical review, current
  status and future potential, Computerized medical imaging and graphics
  31~(4-5) (2007) 198--211.

\bibitem{lavin1998system}
M.~Lavin, M.~Nathan, System and method for managing patient medical records, uS
  Patent 5,772,585 (Jun.~30 1998).

\bibitem{taylor2016medical}
R.~H. Taylor, A.~Menciassi, G.~Fichtinger, P.~Fiorini, P.~Dario, Medical
  robotics and computer-integrated surgery, in: Springer handbook of robotics,
  Springer, 2016, pp. 1657--1684.

\bibitem{litjens2017survey}
G.~Litjens, T.~Kooi, B.~E. Bejnordi, A.~A.~A. Setio, F.~Ciompi, M.~Ghafoorian,
  J.~A. van~der Laak, B.~Van~Ginneken, C.~I. S{\'a}nchez, A survey on deep
  learning in medical image analysis, Medical image analysis 42 (2017) 60--88.

\bibitem{louis20162016}
D.~N. Louis, A.~Perry, G.~Reifenberger, A.~Von~Deimling, D.~Figarella-Branger,
  W.~K. Cavenee, H.~Ohgaki, O.~D. Wiestler, P.~Kleihues, D.~W. Ellison, The
  2016 world health organization classification of tumors of the central
  nervous system: a summary, Acta neuropathologica 131~(6) (2016) 803--820.

\bibitem{baid2021rsna}
U.~Baid, S.~Ghodasara, S.~Mohan, M.~Bilello, E.~Calabrese, E.~Colak,
  K.~Farahani, J.~Kalpathy-Cramer, F.~C. Kitamura, S.~Pati, et~al., The
  rsna-asnr-miccai brats 2021 benchmark on brain tumor segmentation and
  radiogenomic classification, arXiv preprint arXiv:2107.02314.

\bibitem{bakas2017advancing}
S.~Bakas, H.~Akbari, A.~Sotiras, M.~Bilello, M.~Rozycki, J.~S. Kirby, J.~B.
  Freymann, K.~Farahani, C.~Davatzikos, Advancing the cancer genome atlas
  glioma mri collections with expert segmentation labels and radiomic features,
  Scientific data 4~(1) (2017) 1--13.

\bibitem{menze2014multimodal}
B.~H. Menze, A.~Jakab, S.~Bauer, J.~Kalpathy-Cramer, K.~Farahani, J.~Kirby,
  Y.~Burren, N.~Porz, J.~Slotboom, R.~Wiest, et~al., The multimodal brain tumor
  image segmentation benchmark (brats), IEEE transactions on medical imaging
  34~(10) (2014) 1993--2024.

\bibitem{ghaffari2019automated}
M.~Ghaffari, A.~Sowmya, R.~Oliver, Automated brain tumor segmentation using
  multimodal brain scans: a survey based on models submitted to the brats
  2012--2018 challenges, IEEE reviews in biomedical engineering 13 (2019)
  156--168.

\bibitem{kapoor2017survey}
L.~Kapoor, S.~Thakur, A survey on brain tumor detection using image processing
  techniques, in: 2017 7th international conference on cloud computing, data
  science \& engineering-confluence, IEEE, 2017, pp. 582--585.

\bibitem{hameurlaine2019survey}
M.~Hameurlaine, A.~Moussaoui, Survey of brain tumor segmentation techniques on
  magnetic resonance imaging, Nano Biomedicine and Engineering 11~(2) (2019)
  178--191.

\bibitem{gordillo2013state}
N.~Gordillo, E.~Montseny, P.~Sobrevilla, State of the art survey on mri brain
  tumor segmentation, Magnetic resonance imaging 31~(8) (2013) 1426--1438.

\bibitem{liu2014survey}
J.~Liu, M.~Li, J.~Wang, F.~Wu, T.~Liu, Y.~Pan, A survey of mri-based brain
  tumor segmentation methods, Tsinghua Science and Technology 19~(6) (2014)
  578--595.

\bibitem{nalepa2019data}
J.~Nalepa, M.~Marcinkiewicz, M.~Kawulok, Data augmentation for brain-tumor
  segmentation: a review, Frontiers in computational neuroscience 13 (2019) 83.

\bibitem{bernal2019deep}
J.~Bernal, K.~Kushibar, D.~S. Asfaw, S.~Valverde, A.~Oliver, R.~Mart{\'\i},
  X.~Llad{\'o}, Deep convolutional neural networks for brain image analysis on
  magnetic resonance imaging: a review, Artificial intelligence in medicine 95
  (2019) 64--81.

\bibitem{akkus2017deep}
Z.~Akkus, A.~Galimzianova, A.~Hoogi, D.~L. Rubin, B.~J. Erickson, Deep learning
  for brain mri segmentation: state of the art and future directions, Journal
  of digital imaging 30~(4) (2017) 449--459.

\bibitem{esteva2019guide}
A.~Esteva, A.~Robicquet, B.~Ramsundar, V.~Kuleshov, M.~DePristo, K.~Chou,
  C.~Cui, G.~Corrado, S.~Thrun, J.~Dean, A guide to deep learning in
  healthcare, Nature medicine 25~(1) (2019) 24--29.

\bibitem{liu2020deep}
L.~Liu, W.~Ouyang, X.~Wang, P.~Fieguth, J.~Chen, X.~Liu, M.~Pietik{\"a}inen,
  Deep learning for generic object detection: A survey, International journal
  of computer vision 128~(2) (2020) 261--318.

\bibitem{lecun2015deep}
Y.~LeCun, Y.~Bengio, G.~Hinton, Deep learning, nature 521~(7553) (2015) 436.

\bibitem{gu2018recent}
J.~Gu, Z.~Wang, J.~Kuen, L.~Ma, A.~Shahroudy, B.~Shuai, T.~Liu, X.~Wang,
  G.~Wang, J.~Cai, et~al., Recent advances in convolutional neural networks,
  Pattern Recognition 77 (2018) 354--377.

\bibitem{bernal2018deep}
J.~Bernal, K.~Kushibar, D.~S. Asfaw, S.~Valverde, A.~Oliver, R.~Mart{\'\i},
  X.~Llad{\'o}, Deep convolutional neural networks for brain image analysis on
  magnetic resonance imaging: a review, Artificial intelligence in medicine.

\bibitem{Esteva2019}
A.~Esteva, A.~Robicquet, B.~Ramsundar, V.~Kuleshov, M.~DePristo, K.~Chou,
  C.~Cui, G.~Corrado, S.~Thrun, J.~Dean, A guide to deep learning in
  healthcare, Nature Medicine 25~(1) (2019) 24--29.

\bibitem{goodfellow2016deep}
I.~Goodfellow, Y.~Bengio, A.~Courville, Deep learning, 2016.

\bibitem{lin2018resnet}
H.~Lin, S.~Jegelka, Resnet with one-neuron hidden layers is a universal
  approximator, Advances in Neural Information Processing Systems 31 (2018)
  6169--6178.

\bibitem{yarotsky2017error}
D.~Yarotsky, Error bounds for approximations with deep relu networks, Neural
  Networks 94 (2017) 103--114.

\bibitem{liu2016ssd}
W.~Liu, D.~Anguelov, D.~Erhan, C.~Szegedy, S.~Reed, C.-Y. Fu, A.~C. Berg, Ssd:
  Single shot multibox detector, in: European conference on computer vision,
  Springer, 2016, pp. 21--37.

\bibitem{chen2021understanding}
Y.~Chen, J.~Joo, Understanding and mitigating annotation bias in facial
  expression recognition, in: Proceedings of the IEEE/CVF International
  Conference on Computer Vision, 2021, pp. 14980--14991.

\bibitem{bulo2017loss}
S.~R. Bulo, G.~Neuhold, P.~Kontschieder, Loss max-pooling for semantic image
  segmentation, in: 2017 IEEE Conference on Computer Vision and Pattern
  Recognition (CVPR), IEEE, 2017, pp. 7082--7091.

\bibitem{zhu1997computerized}
Y.~Zhu, Z.~Yan, Computerized tumor boundary detection using a hopfield neural
  network, IEEE transactions on medical imaging 16~(1) (1997) 55--67.

\bibitem{clark1998automatic}
M.~C. Clark, L.~O. Hall, D.~B. Goldgof, R.~Velthuizen, F.~R. Murtagh, M.~S.
  Silbiger, Automatic tumor segmentation using knowledge-based techniques, IEEE
  transactions on medical imaging 17~(2) (1998) 187--201.

\bibitem{kaus1999segmentation}
M.~Kaus, S.~K. Warfield, A.~Nabavi, E.~Chatzidakis, P.~M. Black, F.~A. Jolesz,
  R.~Kikinis, Segmentation of meningiomas and low grade gliomas in mri, in:
  International conference on medical image computing and computer-assisted
  intervention, Springer, 1999, pp. 1--10.

\bibitem{prastawa2004brain}
M.~Prastawa, E.~Bullitt, S.~Ho, G.~Gerig, A brain tumor segmentation framework
  based on outlier detection, Medical image analysis 8~(3) (2004) 275--283.

\bibitem{corso2008efficient}
J.~J. Corso, E.~Sharon, S.~Dube, S.~El-Saden, U.~Sinha, A.~Yuille, Efficient
  multilevel brain tumor segmentation with integrated bayesian model
  classification, IEEE transactions on medical imaging 27~(5) (2008) 629--640.

\bibitem{wels2008discriminative}
M.~Wels, G.~Carneiro, A.~Aplas, M.~Huber, J.~Hornegger, D.~Comaniciu, A
  discriminative model-constrained graph cuts approach to fully automated
  pediatric brain tumor segmentation in 3-d mri, in: International Conference
  on Medical Image Computing and Computer-Assisted Intervention, Springer,
  2008, pp. 67--75.

\bibitem{menze2010generative}
B.~H. Menze, K.~Van~Leemput, D.~Lashkari, M.-A. Weber, N.~Ayache, P.~Golland, A
  generative model for brain tumor segmentation in multi-modal images, in:
  International Conference on Medical Image Computing and Computer-Assisted
  Intervention, Springer, 2010, pp. 151--159.

\bibitem{krizhevsky2012imagenet}
A.~Krizhevsky, I.~Sutskever, G.~E. Hinton, Imagenet classification with deep
  convolutional neural networks, Advances in neural information processing
  systems 25 (2012) 1097--1105.

\bibitem{zikic2014segmentation}
D.~Zikic, Y.~Ioannou, M.~Brown, A.~Criminisi, Segmentation of brain tumor
  tissues with convolutional neural networks, Proceedings MICCAI-BRATS 36
  (2014) 36--39.

\bibitem{havaei2017brain}
M.~Havaei, A.~Davy, D.~Warde-Farley, A.~Biard, A.~Courville, Y.~Bengio, C.~Pal,
  P.-M. Jodoin, H.~Larochelle, Brain tumor segmentation with deep neural
  networks, Medical image analysis 35 (2017) 18--31.

\bibitem{pereira2016brain}
S.~Pereira, A.~Pinto, V.~Alves, C.~A. Silva, Brain tumor segmentation using
  convolutional neural networks in mri images, IEEE transactions on medical
  imaging 35~(5) (2016) 1240--1251.

\bibitem{long2015fully}
J.~Long, E.~Shelhamer, T.~Darrell, Fully convolutional networks for semantic
  segmentation, in: Proceedings of the IEEE conference on computer vision and
  pattern recognition, 2015, pp. 3431--3440.

\bibitem{ronneberger2015u}
O.~Ronneberger, P.~Fischer, T.~Brox, U-net: Convolutional networks for
  biomedical image segmentation, in: International Conference on Medical image
  computing and computer-assisted intervention, Springer, 2015, pp. 234--241.

\bibitem{isensee2018no}
F.~Isensee, P.~Kickingereder, W.~Wick, M.~Bendszus, K.~H. Maier-Hein, No
  new-net, in: International MICCAI Brainlesion Workshop, Springer, 2018, pp.
  234--244.

\bibitem{zhao2018deep}
X.~Zhao, Y.~Wu, G.~Song, Z.~Li, Y.~Zhang, Y.~Fan, A deep learning model
  integrating fcnns and crfs for brain tumor segmentation, Medical image
  analysis 43 (2018) 98--111.

\bibitem{jiang2019two}
Z.~Jiang, C.~Ding, M.~Liu, D.~Tao, Two-stage cascaded u-net: 1st place solution
  to brats challenge 2019 segmentation task, in: International MICCAI
  Brainlesion Workshop, Springer, 2019, pp. 231--241.

\bibitem{kamnitsas2017ensembles}
K.~Kamnitsas, W.~Bai, E.~Ferrante, S.~McDonagh, M.~Sinclair, N.~Pawlowski,
  M.~Rajchl, M.~Lee, B.~Kainz, D.~Rueckert, et~al., Ensembles of multiple
  models and architectures for robust brain tumour segmentation, in:
  International MICCAI brainlesion workshop, Springer, 2017, pp. 450--462.

\bibitem{wang2017automatic}
G.~Wang, W.~Li, S.~Ourselin, T.~Vercauteren, Automatic brain tumor segmentation
  using cascaded anisotropic convolutional neural networks, in: International
  MICCAI brainlesion workshop, Springer, 2017, pp. 178--190.

\bibitem{myronenko20183d}
A.~Myronenko, 3d mri brain tumor segmentation using autoencoder regularization,
  in: International MICCAI Brainlesion Workshop, Springer, 2018, pp. 311--320.

\bibitem{zhou2020one}
C.~Zhou, C.~Ding, X.~Wang, Z.~Lu, D.~Tao, One-pass multi-task networks with
  cross-task guided attention for brain tumor segmentation, IEEE Transactions
  on Image Processing 29 (2020) 4516--4529.

\bibitem{sudre2017generalised}
C.~H. Sudre, W.~Li, T.~Vercauteren, S.~Ourselin, M.~J. Cardoso, Generalised
  dice overlap as a deep learning loss function for highly unbalanced
  segmentations, in: Deep learning in medical image analysis and multimodal
  learning for clinical decision support, Springer, 2017, pp. 240--248.

\bibitem{zhang2021cross}
D.~Zhang, G.~Huang, Q.~Zhang, J.~Han, J.~Han, Y.~Yu, Cross-modality deep
  feature learning for brain tumor segmentation, Pattern Recognition 110 (2021)
  107562.

\bibitem{9399263}
T.~Zhou, S.~Canu, P.~Vera, S.~Ruan, Latent correlation representation learning
  for brain tumor segmentation with missing mri modalities, IEEE Transactions
  on Image Processing 30 (2021) 4263--4274.
\newblock \href {http://dx.doi.org/10.1109/TIP.2021.3070752}
  {\path{doi:10.1109/TIP.2021.3070752}}.

\bibitem{de2015deep}
A.~de~Brebisson, G.~Montana, Deep neural networks for anatomical brain
  segmentation, in: Proceedings of the IEEE Conference on Computer Vision and
  Pattern Recognition Workshops, 2015, pp. 20--28.

\bibitem{patenaude2011bayesian}
B.~Patenaude, S.~M. Smith, D.~N. Kennedy, M.~Jenkinson, A bayesian model of
  shape and appearance for subcortical brain segmentation, Neuroimage 56~(3)
  (2011) 907--922.

\bibitem{dou2015automatic}
Q.~Dou, H.~Chen, L.~Yu, L.~Shi, D.~Wang, V.~C. Mok, P.~A. Heng, Automatic
  cerebral microbleeds detection from mr images via independent subspace
  analysis based hierarchical features, in: Engineering in Medicine and Biology
  Society (EMBC), 2015 37th Annual International Conference of the IEEE, IEEE,
  2015, pp. 7933--7936.

\bibitem{dou2016automatic}
Q.~Dou, H.~Chen, L.~Yu, L.~Zhao, J.~Qin, D.~Wang, V.~C. Mok, L.~Shi, P.-A.
  Heng, Automatic detection of cerebral microbleeds from mr images via 3d
  convolutional neural networks, IEEE transactions on medical imaging 35~(5)
  (2016) 1182--1195.

\bibitem{ghafoorian2017deep}
M.~Ghafoorian, N.~Karssemeijer, T.~Heskes, M.~Bergkamp, J.~Wissink, J.~Obels,
  K.~Keizer, F.-E. de~Leeuw, B.~van Ginneken, E.~Marchiori, et~al., Deep
  multi-scale location-aware 3d convolutional neural networks for automated
  detection of lacunes of presumed vascular origin, NeuroImage: Clinical 14
  (2017) 391--399.

\bibitem{suk2016state}
H.-I. Suk, C.-Y. Wee, S.-W. Lee, D.~Shen, State-space model with deep learning
  for functional dynamics estimation in resting-state fmri, NeuroImage 129
  (2016) 292--307.

\bibitem{suk2016deep}
H.-I. Suk, D.~Shen, Deep ensemble sparse regression network for alzheimer’s
  disease diagnosis, in: International Workshop on Machine Learning in Medical
  Imaging, Springer, 2016, pp. 113--121.

\bibitem{pinaya2016using}
W.~H. Pinaya, A.~Gadelha, O.~M. Doyle, C.~Noto, A.~Zugman, Q.~Cordeiro, A.~P.
  Jackowski, R.~A. Bressan, J.~R. Sato, Using deep belief network modelling to
  characterize differences in brain morphometry in schizophrenia, Scientific
  reports 6 (2016) 38897.

\bibitem{yoo2016deep}
Y.~Yoo, L.~W. Tang, T.~Brosch, D.~K. Li, L.~Metz, A.~Traboulsee, R.~Tam, Deep
  learning of brain lesion patterns for predicting future disease activity in
  patients with early symptoms of multiple sclerosis, in: Deep Learning and
  Data Labeling for Medical Applications, Springer, 2016, pp. 86--94.

\bibitem{van2017deep}
H.~K. van~der Burgh, R.~Schmidt, H.-J. Westeneng, M.~A. de~Reus, L.~H. van~den
  Berg, M.~P. van~den Heuvel, Deep learning predictions of survival based on
  mri in amyotrophic lateral sclerosis, NeuroImage: Clinical 13 (2017)
  361--369.

\bibitem{li2017brain}
X.~Li, X.~Zhang, Z.~Luo, Brain tumor segmentation via 3d fully dilated
  convolutional networks, in: Multimodal Brain Tumor Segmentation Benchmark,
  Brain-lesion Workshop, MICCAI, Vol.~9, 2017, p. 2017.

\bibitem{lopez2017dilated}
M.~M. Lopez, J.~Ventura, Dilated convolutions for brain tumor segmentation in
  mri scans, in: International MICCAI Brainlesion Workshop, Springer, 2017, pp.
  253--262.

\bibitem{zhao2017automatic}
L.~Zhao, Automatic brain tumor segmentation with 3d deconvolution network with
  dilated inception block, MICCAI BraTS (2017) 316--320.

\bibitem{islam2019brain}
M.~Islam, V.~Vibashan, V.~J.~M. Jose, N.~Wijethilake, U.~Utkarsh, H.~Ren, Brain
  tumor segmentation and survival prediction using 3d attention unet, in:
  International MICCAI Brainlesion Workshop, Springer, 2019, pp. 262--272.

\bibitem{wang2019global}
H.~Wang, G.~Wang, Z.~Liu, S.~Zhang, Global and local multi-scale feature fusion
  enhancement for brain tumor segmentation and pancreas segmentation, in:
  International MICCAI Brainlesion Workshop, Springer, 2019, pp. 80--88.

\bibitem{liu2020brain}
C.~Liu, W.~Ding, L.~Li, Z.~Zhang, C.~Pei, L.~Huang, X.~Zhuang, Brain tumor
  segmentation network using attention-based fusion and spatial relationship
  constraint, arXiv preprint arXiv:2010.15647.

\bibitem{zhou2020multi}
T.~Zhou, S.~Ruan, Y.~Guo, S.~Canu, A multi-modality fusion network based on
  attention mechanism for brain tumor segmentation, in: 2020 IEEE 17th
  international symposium on biomedical imaging (ISBI), IEEE, 2020, pp.
  377--380.

\bibitem{andermatt2017multi}
S.~Andermatt, S.~Pezold, P.~Cattin, Multi-dimensional gated recurrent units for
  brain tumor segmentation, in: International MICCAI BraTS Challenge.
  Pre-Conference Proceedings, 2017, pp. 15--19.

\bibitem{brugger2019partially}
R.~Br{\"u}gger, C.~F. Baumgartner, E.~Konukoglu, A partially reversible u-net
  for memory-efficient volumetric image segmentation, in: International
  conference on medical image computing and computer-assisted intervention,
  Springer, 2019, pp. 429--437.

\bibitem{chen20193d}
C.~Chen, X.~Liu, M.~Ding, J.~Zheng, J.~Li, 3d dilated multi-fiber network for
  real-time brain tumor segmentation in mri, in: International Conference on
  Medical Image Computing and Computer-Assisted Intervention, Springer, 2019,
  pp. 184--192.

\bibitem{cheng2019memory}
X.~Cheng, Z.~Jiang, Q.~Sun, J.~Zhang, Memory-efficient cascade 3d u-net for
  brain tumor segmentation, in: International MICCAI Brainlesion Workshop,
  Springer, 2019, pp. 242--253.

\bibitem{pendse2020memory}
M.~Pendse, V.~Thangarasa, V.~Chiley, R.~Holmdahl, J.~Hestness, D.~DeCoste,
  Memory efficient 3d u-net with reversible mobile inverted bottlenecks for
  brain tumor segmentation, in: International MICCAI Brainlesion Workshop,
  Springer, 2020, pp. 388--397.

\bibitem{zhao2016multiscale}
L.~Zhao, K.~Jia, Multiscale cnns for brain tumor segmentation and diagnosis,
  Computational and mathematical methods in medicine 2016.

\bibitem{shen2017efficient}
H.~Shen, J.~Zhang, W.~Zheng, Efficient symmetry-driven fully convolutional
  network for multimodal brain tumor segmentation, in: 2017 IEEE International
  Conference on Image Processing (ICIP), IEEE, 2017, pp. 3864--3868.

\bibitem{castillo2017volumetric}
L.~S. Castillo, L.~A. Daza, L.~C. Rivera, P.~Arbel{\'a}ez, Volumetric
  multimodality neural network for brain tumor segmentation, in: 13th
  international conference on medical information processing and analysis, Vol.
  10572, International Society for Optics and Photonics, 2017, p. 105720E.

\bibitem{jungo2017towards}
A.~Jungo, R.~McKinley, R.~Meier, U.~Knecht, L.~Vera, J.~P{\'e}rez-Beteta,
  D.~Molina-Garc{\'\i}a, V.~M. P{\'e}rez-Garc{\'\i}a, R.~Wiest, M.~Reyes,
  Towards uncertainty-assisted brain tumor segmentation and survival
  prediction, in: International MICCAI Brainlesion Workshop, Springer, 2017,
  pp. 474--485.

\bibitem{shaikh2017brain}
M.~Shaikh, G.~Anand, G.~Acharya, A.~Amrutkar, V.~Alex, G.~Krishnamurthi, Brain
  tumor segmentation using dense fully convolutional neural network, in:
  International MICCAI brainlesion workshop, Springer, 2017, pp. 309--319.

\bibitem{zhou2020afpnet}
Z.~Zhou, Z.~He, Y.~Jia, Afpnet: A 3d fully convolutional neural network with
  atrous-convolution feature pyramid for brain tumor segmentation via mri
  images, Neurocomputing 402 (2020) 235--244.

\bibitem{zhou2021latent}
T.~Zhou, S.~Canu, P.~Vera, S.~Ruan, Latent correlation representation learning
  for brain tumor segmentation with missing mri modalities, IEEE Transactions
  on Image Processing 30 (2021) 4263--4274.

\bibitem{dvovrak2015local}
P.~Dvo{\v{r}}{\'a}k, B.~Menze, Local structure prediction with convolutional
  neural networks for multimodal brain tumor segmentation, in: International
  MICCAI workshop on medical computer vision, Springer, 2015, pp. 59--71.

\bibitem{rao2015brain}
V.~Rao, M.~S. Sarabi, A.~Jaiswal, Brain tumor segmentation with deep learning,
  MICCAI Multimodal Brain Tumor Segmentation Challenge (BraTS) 59.

\bibitem{simonyan2014very}
K.~Simonyan, A.~Zisserman, Very deep convolutional networks for large-scale
  image recognition, arXiv preprint arXiv:1409.1556.

\bibitem{casamitjana20163d}
A.~Casamitjana, S.~Puch, A.~Aduriz, E.~Sayrol, V.~Vilaplana, 3d convolutional
  networks for brain tumor segmentation, Proceedings of the MICCAI Challenge on
  Multimodal Brain Tumor Image Segmentation (BRATS) (2016) 65--68.

\bibitem{jesson2017brain}
A.~Jesson, T.~Arbel, Brain tumor segmentation using a 3d fcn with multi-scale
  loss, in: International MICCAI Brainlesion Workshop, Springer, 2017, pp.
  392--402.

\bibitem{chang2016fully}
P.~D. Chang, Fully convolutional deep residual neural networks for brain tumor
  segmentation, in: International workshop on Brainlesion: Glioma, multiple
  sclerosis, stroke and traumatic brain injuries, Springer, 2016, pp. 108--118.

\bibitem{he2016deep}
K.~He, X.~Zhang, S.~Ren, J.~Sun, Deep residual learning for image recognition,
  in: Proceedings of the IEEE conference on computer vision and pattern
  recognition, 2016, pp. 770--778.

\bibitem{ghaffari2020brain}
M.~Ghaffari, A.~Sowmya, R.~Oliver, Brain tumour segmentation using cascaded 3d
  densely-connected u-net (2020).
\newblock \href {http://arxiv.org/abs/2009.07563} {\path{arXiv:2009.07563}}.

\bibitem{wang2020modality}
Y.~Wang, Y.~Zhang, F.~Hou, Y.~Liu, J.~Tian, C.~Zhong, Y.~Zhang, Z.~He,
  Modality-pairing learning for brain tumor segmentation, arXiv preprint
  arXiv:2010.09277.

\bibitem{DBLP:journals/cbm/ZhouHSDC20}
Z.~Zhou, Z.~He, M.~Shi, J.~Du, D.~Chen,
  \href{https://doi.org/10.1016/j.compbiomed.2020.103766}{3d dense connectivity
  network with atrous convolutional feature pyramid for brain tumor
  segmentation in magnetic resonance imaging of human heads}, Comput. Biol.
  Medicine 121 (2020) 103766.
\newblock \href {http://dx.doi.org/10.1016/j.compbiomed.2020.103766}
  {\path{doi:10.1016/j.compbiomed.2020.103766}}.
\newline\urlprefix\url{https://doi.org/10.1016/j.compbiomed.2020.103766}

\bibitem{huang2017densely}
G.~Huang, Z.~Liu, L.~Van Der~Maaten, K.~Q. Weinberger, Densely connected
  convolutional networks, in: Proceedings of the IEEE conference on computer
  vision and pattern recognition, 2017, pp. 4700--4708.

\bibitem{yu2017dilated}
F.~Yu, V.~Koltun, T.~Funkhouser, Dilated residual networks, in: Proceedings of
  the IEEE conference on computer vision and pattern recognition, 2017, pp.
  472--480.

\bibitem{choudhury2018segmentation}
A.~R. Choudhury, R.~Vanguri, S.~R. Jambawalikar, P.~Kumar, Segmentation of
  brain tumors using deeplabv3+, in: International MICCAI Brainlesion Workshop,
  Springer, 2018, pp. 154--167.

\bibitem{andermatt2016multi}
S.~Andermatt, S.~Pezold, P.~Cattin, Multi-dimensional gated recurrent units for
  the segmentation of biomedical 3d-data, in: Deep learning and data labeling
  for medical applications, Springer, 2016, pp. 142--151.

\bibitem{gomez2017reversible}
A.~N. Gomez, M.~Ren, R.~Urtasun, R.~B. Grosse, The reversible residual network:
  Backpropagation without storing activations, in: Proceedings of the 31st
  International Conference on Neural Information Processing Systems, 2017, pp.
  2211--2221.

\bibitem{sandler2018mobilenetv2}
M.~Sandler, A.~Howard, M.~Zhu, A.~Zhmoginov, L.-C. Chen, Mobilenetv2: Inverted
  residuals and linear bottlenecks, in: Proceedings of the IEEE conference on
  computer vision and pattern recognition, 2018, pp. 4510--4520.

\bibitem{tan2019efficientnet}
M.~Tan, Q.~Le, Efficientnet: Rethinking model scaling for convolutional neural
  networks, in: International Conference on Machine Learning, PMLR, 2019, pp.
  6105--6114.

\bibitem{nuechterlein20183d}
N.~Nuechterlein, S.~Mehta, 3d-espnet with pyramidal refinement for volumetric
  brain tumor image segmentation, in: International MICCAI Brainlesion
  Workshop, Springer, 2018, pp. 245--253.

\bibitem{urban2014multi}
G.~Urban, M.~Bendszus, F.~Hamprecht, J.~Kleesiek, Multi-modal brain tumor
  segmentation using deep convolutional neural networks, MICCAI BraTS (brain
  tumor segmentation) challenge. Proceedings, winning contribution (2014)
  31--35.

\bibitem{akil2020fully}
M.~Akil, R.~Saouli, R.~Kachouri, et~al., Fully automatic brain tumor
  segmentation with deep learning-based selective attention using overlapping
  patches and multi-class weighted cross-entropy, Medical image analysis 63
  (2020) 101692.

\bibitem{kamnitsas2017efficient}
K.~Kamnitsas, C.~Ledig, V.~F. Newcombe, J.~P. Simpson, A.~D. Kane, D.~K. Menon,
  D.~Rueckert, B.~Glocker, Efficient multi-scale 3d cnn with fully connected
  crf for accurate brain lesion segmentation, Medical image analysis 36 (2017)
  61--78.

\bibitem{shen2017boundary}
H.~Shen, R.~Wang, J.~Zhang, S.~J. McKenna, Boundary-aware fully convolutional
  network for brain tumor segmentation, in: International Conference on Medical
  Image Computing and Computer-Assisted Intervention, Springer, 2017, pp.
  433--441.

\bibitem{brosch2016deep}
T.~Brosch, L.~Y. Tang, Y.~Yoo, D.~K. Li, A.~Traboulsee, R.~Tam, Deep 3d
  convolutional encoder networks with shortcuts for multiscale feature
  integration applied to multiple sclerosis lesion segmentation, IEEE
  transactions on medical imaging 35~(5) (2016) 1229--1239.

\bibitem{isensee2017brain}
F.~Isensee, P.~Kickingereder, W.~Wick, M.~Bendszus, K.~H. Maier-Hein, Brain
  tumor segmentation and radiomics survival prediction: Contribution to the
  brats 2017 challenge, in: International MICCAI Brainlesion Workshop,
  Springer, 2017, pp. 287--297.

\bibitem{dong2017automatic}
H.~Dong, G.~Yang, F.~Liu, Y.~Mo, Y.~Guo, Automatic brain tumor detection and
  segmentation using u-net based fully convolutional networks, in: annual
  conference on medical image understanding and analysis, Springer, 2017, pp.
  506--517.

\bibitem{milletari2016v}
F.~Milletari, N.~Navab, S.-A. Ahmadi, V-net: Fully convolutional neural
  networks for volumetric medical image segmentation, in: 3D Vision (3DV), 2016
  Fourth International Conference on, IEEE, 2016, pp. 565--571.

\bibitem{beers2017sequential}
A.~Beers, K.~Chang, J.~Brown, E.~Sartor, C.~Mammen, E.~Gerstner, B.~Rosen,
  J.~Kalpathy-Cramer, Sequential 3d u-nets for biologically-informed brain
  tumor segmentation, arXiv preprint arXiv:1709.02967.

\bibitem{chen2017brain}
S.~Chen, C.~Ding, C.~Zhou, Brain tumor segmentation with label distribution
  learning and multi-level feature representation, 2017 International MICCAI
  BraTS Challenge.

\bibitem{chen2019dual}
S.~Chen, C.~Ding, M.~Liu, Dual-force convolutional neural networks for accurate
  brain tumor segmentation, Pattern Recognition 88 (2019) 90--100.

\bibitem{pawar2017residual}
K.~Pawar, Z.~Chen, N.~J. Shah, G.~Egan, Residual encoder and convolutional
  decoder neural network for glioma segmentation, in: International MICCAI
  Brainlesion Workshop, Springer, 2017, pp. 263--273.

\bibitem{chen2018s3d}
W.~Chen, B.~Liu, S.~Peng, J.~Sun, X.~Qiao, S3d-unet: separable 3d u-net for
  brain tumor segmentation, in: International MICCAI Brainlesion Workshop,
  Springer, 2018, pp. 358--368.

\bibitem{fang2018three}
L.~Fang, H.~He, Three pathways u-net for brain tumor segmentation, in:
  Pre-conference proceedings of the 7th medical image computing and
  computer-assisted interventions (MICCAI) BraTS Challenge, Vol. 2018, 2018,
  pp. 119--126.

\bibitem{hua2018multimodal}
R.~Hua, Q.~Huo, Y.~Gao, Y.~Sun, F.~Shi, Multimodal brain tumor segmentation
  using cascaded v-nets, in: International MICCAI Brainlesion Workshop,
  Springer, 2018, pp. 49--60.

\bibitem{li2018fused}
X.~Li, Fused u-net for brain tumor segmentation based on multimodal mr images,
  International MICCAI Brain Tumor Segmentation (BraTS) challenge (2018)
  290--297.

\bibitem{zhao2019bag}
Y.-X. Zhao, Y.-M. Zhang, C.-L. Liu, Bag of tricks for 3d mri brain tumor
  segmentation, in: International MICCAI Brainlesion Workshop, Springer, 2019,
  pp. 210--220.

\bibitem{Yuan2020AutomaticBT}
Y.~Yuan, Automatic brain tumor segmentation with scale attention network, in:
  BrainLes@MICCAI, 2020.

\bibitem{henry2020brain}
T.~Henry, A.~Carre, M.~Lerousseau, T.~Estienne, C.~Robert, N.~Paragios,
  E.~Deutsch, Brain tumor segmentation with self-ensembled, deeply-supervised
  3d u-net neural networks: a brats 2020 challenge solution, arXiv preprint
  arXiv:2011.01045.

\bibitem{bae2019resource}
W.~Bae, S.~Lee, Y.~Lee, B.~Park, M.~Chung, K.-H. Jung, Resource optimized
  neural architecture search for 3d medical image segmentation, in:
  International Conference on Medical Image Computing and Computer-Assisted
  Intervention, Springer, 2019, pp. 228--236.

\bibitem{kim2019scalable}
S.~Kim, I.~Kim, S.~Lim, W.~Baek, C.~Kim, H.~Cho, B.~Yoon, T.~Kim, Scalable
  neural architecture search for 3d medical image segmentation, in:
  International Conference on Medical Image Computing and Computer-Assisted
  Intervention, Springer, 2019, pp. 220--228.

\bibitem{zhou2019unet++}
Z.~Zhou, M.~M.~R. Siddiquee, N.~Tajbakhsh, J.~Liang, Unet++: Redesigning skip
  connections to exploit multiscale features in image segmentation, IEEE
  transactions on medical imaging 39~(6) (2019) 1856--1867.

\bibitem{zhu2019v}
Z.~Zhu, C.~Liu, D.~Yang, A.~Yuille, D.~Xu, V-nas: Neural architecture search
  for volumetric medical image segmentation, in: 2019 International Conference
  on 3D Vision (3DV), IEEE, 2019, pp. 240--248.

\bibitem{dong2018imbalanced}
Q.~Dong, S.~Gong, X.~Zhu, Imbalanced deep learning by minority class
  incremental rectification, IEEE transactions on pattern analysis and machine
  intelligence 41~(6) (2018) 1367--1381.

\bibitem{johnson2019survey}
J.~M. Johnson, T.~M. Khoshgoftaar, Survey on deep learning with class
  imbalance, Journal of Big Data 6~(1) (2019) 1--54.

\bibitem{jia2020h2nf}
H.~Jia, W.~Cai, H.~Huang, Y.~Xia, H2nf-net for brain tumor segmentation using
  multimodal mr imaging: 2nd place solution to brats challenge 2020
  segmentation task, in: BrainLes@ MICCAI (2), 2020.

\bibitem{sun2019deep}
K.~Sun, B.~Xiao, D.~Liu, J.~Wang, Deep high-resolution representation learning
  for human pose estimation, in: CVPR, 2019.

\bibitem{li2019multi}
X.~Li, G.~Luo, K.~Wang, Multi-step cascaded networks for brain tumor
  segmentation, in: International MICCAI Brainlesion Workshop, Springer, 2019,
  pp. 163--173.

\bibitem{vu2019tunet}
M.~H. Vu, T.~Nyholm, T.~L{\"o}fstedt, Tunet: End-to-end hierarchical brain
  tumor segmentation using cascaded networks, in: International MICCAI
  Brainlesion Workshop, Springer, 2019, pp. 174--186.

\bibitem{liu2020automatic}
Z.~Liu, D.~Gu, Y.~Zhang, X.~Cao, Z.~Xue, Automatic segmentation of non-tumor
  tissues in glioma mr brain images using deformable registration with partial
  convolutional networks, in: International MICCAI Brainlesion Workshop,
  Springer, 2020, pp. 41--50.

\bibitem{cirillo2020vox2vox}
M.~D. Cirillo, D.~Abramian, A.~Eklund, Vox2vox: 3d-gan for brain tumour
  segmentation, arXiv preprint arXiv:2003.13653.

\bibitem{chen2020brain}
H.~Chen, Z.~Qin, Y.~Ding, L.~Tian, Z.~Qin, Brain tumor segmentation with deep
  convolutional symmetric neural network, Neurocomputing 392 (2020) 305--313.

\bibitem{kamnitsas2016deepmedic}
K.~Kamnitsas, E.~Ferrante, S.~Parisot, C.~Ledig, A.~V. Nori, A.~Criminisi,
  D.~Rueckert, B.~Glocker, Deepmedic for brain tumor segmentation, in:
  International workshop on Brainlesion: Glioma, multiple sclerosis, stroke and
  traumatic brain injuries, Springer, 2016, pp. 138--149.

\bibitem{kao2018brain}
P.-Y. Kao, T.~Ngo, A.~Zhang, J.~W. Chen, B.~Manjunath, Brain tumor segmentation
  and tractographic feature extraction from structural mr images for overall
  survival prediction, in: International MICCAI Brainlesion Workshop, Springer,
  2018, pp. 128--141.

\bibitem{lachinov2019knowledge}
D.~Lachinov, E.~Shipunova, V.~Turlapov, Knowledge distillation for brain tumor
  segmentation, in: International MICCAI Brainlesion Workshop, Springer, 2019,
  pp. 324--332.

\bibitem{lachinov2018glioma}
D.~Lachinov, E.~Vasiliev, V.~Turlapov, Glioma segmentation with cascaded unet,
  in: International MICCAI Brainlesion Workshop, Springer, 2018, pp. 189--198.

\bibitem{zhao20173d}
X.~Zhao, Y.~Wu, G.~Song, Z.~Li, Y.~Zhang, Y.~Fan, 3d brain tumor segmentation
  through integrating multiple 2d fcnns, in: International MICCAI Brainlesion
  Workshop, Springer, 2017, pp. 191--203.

\bibitem{sundaresan2020brain}
V.~Sundaresan, L.~Griffanti, M.~Jenkinson, Brain tumour segmentation using a
  triplanar ensemble of u-nets on mr images, in: International MICCAI
  Brainlesion Workshop, Springer, 2020, pp. 340--353.

\bibitem{chen2017fully}
L.~Chen, P.~Bentley, D.~Rueckert, Fully automatic acute ischemic lesion
  segmentation in dwi using convolutional neural networks, NeuroImage: Clinical
  15 (2017) 633--643.

\bibitem{noh2015learning}
H.~Noh, S.~Hong, B.~Han, Learning deconvolution network for semantic
  segmentation, in: Proceedings of the IEEE international conference on
  computer vision, 2015, pp. 1520--1528.

\bibitem{hu20173d}
Y.~Hu, Y.~Xia, 3d deep neural network-based brain tumor segmentation using
  multimodality magnetic resonance sequences, in: International MICCAI
  Brainlesion Workshop, Springer, 2017, pp. 423--434.

\bibitem{silva2020multi}
C.~A. Silva, A.~Pinto, S.~Pereira, A.~Lopes, Multi-stage deep layer aggregation
  for brain tumor segmentation, in: International MICCAI Brainlesion Workshop,
  Springer, 2020, pp. 179--188.

\bibitem{zhou2018learning}
C.~Zhou, S.~Chen, C.~Ding, D.~Tao, Learning contextual and attentive
  information for brain tumor segmentation, in: International MICCAI
  brainlesion workshop, Springer, 2018, pp. 497--507.

\bibitem{shen2017multi}
H.~Shen, R.~Wang, J.~Zhang, S.~McKenna, Multi-task fully convolutional network
  for brain tumour segmentation, in: Annual Conference on Medical Image
  Understanding and Analysis, Springer, 2017, pp. 239--248.

\bibitem{nguyen2020enhancing}
H.~T. Nguyen, T.~T. Le, T.~V. Nguyen, N.~T. Nguyen, Enhancing mri brain tumor
  segmentation with an additional classification network, arXiv preprint
  arXiv:2009.12111.

\bibitem{weninger2019multi}
L.~Weninger, Q.~Liu, D.~Merhof, Multi-task learning for brain tumor
  segmentation, in: International MICCAI brainlesion workshop, Springer, 2019,
  pp. 327--337.

\bibitem{iwasawa2020label}
J.~Iwasawa, Y.~Hirano, Y.~Sugawara, Label-efficient multi-task segmentation
  using contrastive learning, arXiv preprint arXiv:2009.11160.

\bibitem{caruana1997multitask}
R.~Caruana, Multitask learning, Machine learning 28~(1) (1997) 41--75.

\bibitem{evgeniou2004regularized}
T.~Evgeniou, M.~Pontil, Regularized multi--task learning, in: Proceedings of
  the tenth ACM SIGKDD international conference on Knowledge discovery and data
  mining, 2004, pp. 109--117.

\bibitem{sener2018multi}
O.~Sener, V.~Koltun, Multi-task learning as multi-objective optimization, in:
  Proceedings of the 32nd International Conference on Neural Information
  Processing Systems, 2018, pp. 525--536.

\bibitem{zhang2021survey}
Y.~Zhang, Q.~Yang, A survey on multi-task learning, IEEE Transactions on
  Knowledge and Data Engineering.

\bibitem{randhawa2016improving}
R.~S. Randhawa, A.~Modi, P.~Jain, P.~Warier, Improving boundary classification
  for brain tumor segmentation and longitudinal disease progression, in:
  International Workshop on Brainlesion: Glioma, Multiple Sclerosis, Stroke and
  Traumatic Brain Injuries, Springer, 2016, pp. 65--74.

\bibitem{pawar2019ensemble}
K.~Pawar, Z.~Chen, N.~J. Shah, G.~F. Egan, An ensemble of 2d convolutional
  neural network for 3d brain tumor segmentation, in: International MICCAI
  Brainlesion Workshop, Springer, 2019, pp. 359--367.

\bibitem{tseng2017joint}
K.-L. Tseng, Y.-L. Lin, W.~Hsu, C.-Y. Huang, Joint sequence learning and
  cross-modality convolution for 3d biomedical segmentation, in: Proceedings of
  the IEEE conference on Computer Vision and Pattern Recognition, 2017, pp.
  6393--6400.

\bibitem{cata2017masked}
M.~Cat{\`a}, A.~Casamitjana~D{\'\i}az, I.~Sanchez~Muriana, M.~Combalia,
  V.~Vilaplana~Besler, Masked v-net: an approach to brain tumor segmentation,
  in: 2017 International MICCAI BraTS Challenge. Pre-conference proceedings,
  2017, pp. 42--49.

\bibitem{zhang2020exploring}
D.~Zhang, G.~Huang, Q.~Zhang, J.~Han, J.~Han, Y.~Wang, Y.~Yu, Exploring task
  structure for brain tumor segmentation from multi-modality mr images, IEEE
  Transactions on Image Processing 29 (2020) 9032--9043.

\bibitem{li2017deep}
Y.~Li, L.~Shen, Deep learning based multimodal brain tumor diagnosis, in:
  International MICCAI Brainlesion Workshop, Springer, 2017, pp. 149--158.

\bibitem{yu20183d}
B.~Yu, L.~Zhou, L.~Wang, J.~Fripp, P.~Bourgeat, 3d cgan based cross-modality mr
  image synthesis for brain tumor segmentation, in: 2018 IEEE 15th
  International Symposium on Biomedical Imaging (ISBI 2018), IEEE, 2018, pp.
  626--630.

\bibitem{zhou2020brain}
T.~Zhou, S.~Canu, P.~Vera, S.~Ruan, Brain tumor segmentation with missing
  modalities via latent multi-source correlation representation, in:
  International Conference on Medical Image Computing and Computer-Assisted
  Intervention, Springer, 2020, pp. 533--541.

\bibitem{yu2021sa}
B.~Yu, L.~Zhou, L.~Wang, W.~Yang, M.~Yang, P.~Bourgeat, J.~Fripp, Sa-lut-nets:
  Learning sample-adaptive intensity lookup tables for brain tumor
  segmentation, IEEE Transactions on Medical Imaging 40~(5) (2021) 1417--1427.

\end{thebibliography}

\end{document}